%% file: paper.tex
#\documentclass[acmsmall,screen]{acmart}
\documentclass[acmsmall]{acmart}

\setcopyright{rightsretained}
\acmPrice{}
\acmDOI{10.1145/3371071}
\acmYear{2020}
\copyrightyear{2020}
\acmJournal{PACMPL}
\acmVolume{4}
\acmNumber{POPL}
\acmArticle{3}
\acmMonth{1}

\usepackage{booktabs}   
\usepackage{subcaption} 

\usepackage{fixltx2e}
\usepackage{alltt}
\usepackage{xcolor}
\usepackage{pifont} 
\usepackage{mathpartir}
\usepackage{tikz}
\usetikzlibrary{shapes.geometric, arrows, shadows, positioning}
\usepackage{textgreek}

\usepackage{paper}

\begin{document}

\title{Dependent Type Systems as Macros}         


\author{Stephen Chang}
\affiliation{
  \institution{Northeastern University}            
  \country{USA}                    
}\affiliation{
  \institution{PLT Group}            
  \country{USA}                    
}
\email{stchang@ccs.neu.edu}          

\author{Michael Ballantyne}
\affiliation{
  \institution{Northeastern University}            
  \country{USA}                    
}
\affiliation{
  \institution{PLT Group}            
  \country{USA}                    
}
\email{mballantyne@ccs.neu.edu}          

\author{Milo Turner}
\affiliation{
  \institution{Northeastern University}            
  \country{USA}                    
}
\email{milo@ccs.neu.edu}          

\author{William J. Bowman}
\affiliation{
  \institution{University of British Columbia}            
  \country{Canada}                    
}
\email{wjb@williamjbowman.com}          


\begin{abstract}
\input{abstract}

\end{abstract}

\ccsdesc[300]{Software and its engineering~Specialized application languages}

\keywords{macros, type systems, dependent types, proof assistants}  

\maketitle

\section{The Trouble with Implementing Dependent Types}\label{sec:intro}
\input{intro}

\section{Creating a Typed Language (STLC) with Racket and \Turntwo}\label{sec:primer}
\input{primer}

\section{Lightweight Dependent Types, for \modu{Video}}\label{sec:video}
\input{video}

\section{A Dependently-Typed Calculus}\label{sec:dep}
\input{dep}



\section{From Calculus to Programming Language: Introducing \modu{Cur}}\label{sec:precur}
\input{precur}

\newcommand{\secsp}{\vspace{-5pt}}
\secsp
\section{Companion DSLs for a Proof Assistant}\label{sec:cur}
\input{cur}

\secsp
\section{Related Work}\label{sec:related}
\input{related}

\secsp
\section{Conclusion}\label{sec:conclusion}

\input{conclusion}

\secsp
\input{acks}

\newpage

\bibliography{depmacros,wjb}




\end{document}

%% file: abstract.tex
We present \Turntwo, a high-level, macros-based metaDSL for building
dependently typed languages.
With it, programmers may rapidly prototype and iterate on the design of new
dependently typed features and extensions.
Or they may create entirely new DSLs whose dependent type ``power'' is tailored
to a specific domain.
Our framework's support of language-oriented programming also makes it suitable
for experimenting with systems of interacting components, e.g., a proof
assistant and its companion DSLs.
This paper explains the implementation details of \Turntwo, as well as how it
may be used to create a wide-variety of dependently typed languages, from a
lightweight one with indexed types, to a full spectrum proof assistant,
complete with a tactic system and extensions for features like sized types and
SMT interaction.

%% file: intro.tex
Dependent types are breaking into the mainstream. For example, Scala supports
path-dependent types~\cite{scalapathdep06, scalapathdep14}, Haskell has
embraced type-level computation~\cite{weirich2017}, and Rust has considered
$\Pi$ types~\shortcite{rust2017b}. Further, interactive languages like F*
are increasingly used to verify critical software such as Firefox's
TLS~\cite{zinzindohoue2017}.

Unfortunately, widespread use remains on hold as language designers
continue exploring the design space, trying to balance the power of dependent
types with their steep learning curve. Worse, because dependent types blur the
distinction between types and runtime values---complicating a language's
implementation as well---evaluating a feature is more difficult and
iterating on its design is extremely time-consuming. Indeed, determining an
ideal ``power-to-weight'' ratio has slowed adoption in
Haskell~\cite{yorgey2012}, and has led to repeated rewrites, and the
abandonment of, Rust dependent type RFCs~\shortcite{rust2017b}. This history
suggests that language designers would benefit from an easier way to build and
iterate on the design of dependently typed features and languages.

We present \Turntwo, a metalanguage for implementing typed---particularly
dependently typed---languages.  \Turntwo supports terse mathematical notation,
modular language feature implementation, and implicit handling of complex type
system implementation patterns, including those for dependent types.  These
capabilities allow language designers to iterate quickly on the design of a new
dependently typed feature by rapidly building modular and extensible prototype
implementations. Or they may drastically reduce the size of their design space
by tailoring the power of a type system to a specific DSL.  \Turntwo's ability
to quickly create languages comes from its use of LISP and Scheme-style
\emph{macros}, which enable reusing a host language's infrastructure when
building new languages.  \Turntwo is implemented in Racket~\cite{manifesto},
this paper's platform of choice, because it exposes more of the compiler for
reuse than its predecessors~\cite{flatt2002, fcdf:macrosworktogether,
  setsofscopes}, and has a language-oriented focus~\cite{lop}.

\Turntwo can also help advance the state-of-the-art for full-spectrum
dependently typed languages like Coq or Agda, which often
have powerful core type theories that are difficult to program directly.  Thus,
users rely on a variety of features layered on top of the core, from extensions
for unification~\cite{mcbride2000:thesis}, termination~\cite{gimenez1995}, or
automation~\cite{hammer2016}, to companion DSLs like tactic
systems~\cite{delahaye2000, mtac, gonthier2010} or pattern
matchers~\cite{coquand1992, norell2007}.
%
Unfortunately, such an ad-hoc system of interacting components can be difficult
to modify, especially when third-party tools are involved. We claim our
macro-based, language-oriented approach creates more extensible, linguistically-integrated
components, and can thus help researchers more easily explore alternate designs.


The main technical contribution of our paper is the design and implementation
of the \Turntwo metalanguage, which is a complete rewrite
of~\citet{macrotypes}'s \Turnstile. \Turnstile showed that typed
languages, instead of being built from scratch, could be implemented by adding
some type checking code to macro definitions, and then reusing the
infrastructure of the macro system---\eg, its implementation of binding,
pattern matching, environments, and transformations---for the rest of the type
checker. Even better, organized in this way, a type checker implementation
closely corresponds to its (algorithmic) specification, and language creators
may exploit this correspondence by coding at the level of mathematical type
rules.  \Turntwo represents a major research leap over its predecessor.
Specifically, we solve the major challenges necessary to implement dependent
types and their accompanying DSLs and extensions (which \Turnstile could not
support), while retaining the original abilities of \Turnstile.
For example, one considerable obstacle was the separation between the macro
expansion phase and a program's runtime phase.
Since dependently typed languages may evaluate expressions while type checking,
checking dependent types with macros requires new macrology design patterns and
abstractions for interleaving expansion, type checking, and evaluation.
The following summarizes our key innovations.
\begin{itemize}

\item \Turntwo demands a radically different API for implementing a
    language's types. It must be straightforward yet expressive enough to represent a
    range of constructs from base types, to binding forms like $\Pi$ types, to
    datatype definition forms for indexed inductive type families.

\item \Turntwo includes an API for defining type-level computation,
    which we dub \emph{normalization by macro expansion}. A programmer writes
    a reduction rule using syntax resembling familiar on-paper notation,
    and \Turntwo generates a macro definition that performs the reduction during
    macro expansion. This allows easily implementing modular type-level
    evaluation.

\item \Turntwo's new type API adds a generic type operation interface,
    enabling modular implementation of features such as error messages, pattern
    matching, and resugaring. This is particularly
    important for implementing tools like tactic systems that inspect
    intermediate type checking steps and construct partial terms.

\item \Turntwo's core type checking infrastructure requires an overhaul,
  specifically with first-class type environments, in order to accommodate
  features like dependent binding structures of the shape $[x:\tau]\ldots$,
  \ie, \emph{telescopes}~\cite{debruijn1991:telescope, mcbride2000:thesis}.

\item Relatedly, \Turntwo's inference-rule syntax is extended
  so that operations over telescopes, or premises with references to telescopes,
  operate as ``fold''s instead of as ``map''s.
\end{itemize}
\noindent{}To evaluate our claim that \Turntwo allows quickly and modularly
iterating on dependently typed languages, and tailoring the power of a type
system, we present a series of examples that range from ``lightweight'' to
``full-spectrum'', though we spend more time on the ``full'' end because it is
more involved. Specifically we show how to create: a video-editing DSL with a
Dependent ML-like type system; a full-spectrum dependently typed calculus \`a
la Martin L\"of; as well as one with inductive datatypes \`a
la~\citet{Dybjer1994}.  To show that our approach scales to a realistic
language, we then turn our core inductive calculus into \Cur, a prototype proof
assistant with: recursive definitions and termination checking; unification and
implicit arguments; and dependent pattern matching.  To show that our system
supports common extensions we create: a tactic system; a metaDSL to implement
the tactic system; a system similar to Ott~\cite{sewell:2007} for writing
language definitions; a library for proving theorems via an SMT solver; and a
library that adds sized types~\cite{hughes96sizedtypes,abel10sized}.  Finally,
to demonstrate that languages created with \Turntwo are realistic to program
with, we used \Cur to work through a graduate-level semester's worth of
examples---roughly volume 1 of the ``Software Foundations''
curriculum~\cite{sf1}.

%% file: primer.tex

\subsection{Interleaving Transformation and Type Checking with Macros}

Macros are special compile-time functions that consume and produce \emph{syntax
  objects}~\cite{dybvig1992}, \ie, enhanced (with source location, binding
structure, etc.) S-expression ASTs. \fullref[center]{fig:tcarch} illustrates
how they are suitable for implementing typed languages because the syntax
transformation and side-condition components of a typical \emph{macro
  definition} mirrors the checking and transformation (\eg, to a lower-level
core language) performed by type checkers (\fullref[left]{fig:tcarch}). More
specifically, a macro definition deconstructs its input via a \emph{syntax
  pattern} (\pat{gray} in this paper),\footnote{To better communicate
  high-level concepts, some code may be stylized, \eg, we may elide unimportant
  details, use abbreviations, \stx{\scriptsize{color}}, or subscripts. For example,
  a \texttt{\defmac} macro is shorthand for
  \texttt{define-syntax} and \texttt{syntax-parse} pattern
  matching. Thus, some code may not run as
  presented, but runnable examples for all code are 
  in our artifact.}
whose shape dictates the macro's usage syntax. This input is eventually
transformed, after checking possible side-condition guards, into a macro's
output, which is typically created with a quasiquotation \emph{syntax template}
(\stx{blue}).

In a language with macros, a \emph{macro expansion} compiler phase repeatedly
rewrites a program's surface syntax according to all macro definitions, until no
macro invocations remain. During expansion, if any of a macro's side-conditions
are not satisfied, compilation fails with an error. By implementing type rules
as these side-conditions, one may implement a type checker as macros. Since
macro definitions are modular components, a macro-based type checker
\emph{interleaves} checking and transformation (\fullref[r]{fig:tcarch}). This
differs from the simple architecture depicted by \fullref[l]{fig:tcarch}, but
it more closely follows the modular nature of a type system's rule
specifications.

%
%
%
\begin{figure}[t]
%
%
\begin{minipage}[c]{0.25\textwidth}
\begin{center}
\tikzstyle{pat} = [rectangle, rounded corners, minimum width=.9cm, minimum height=0.1cm,text justified]
\tikzstyle{stx} = [rectangle, rounded corners, minimum width=.9cm, minimum height=0.1cm,text justified]
\tikzstyle{rect} = [rectangle, rounded corners, minimum width=2cm, minimum height=0.1cm,text justified, draw=black]
\tikzstyle{arrow} = [->,>=stealth, line width=1pt]
\tikzstyle{line} = [-,>=stealth, line width=1pt]

\begin{tikzpicture}[auto, node distance=2cm,>=latex']
\node (Start) [pat] {\codefontsize \pat{surface lang stx}};

\node(Typecheck) [rect, below of=Start, node distance=.8cm] {
\small type check
};

\node (Transform) [rect, below of=Typecheck, node distance=.8cm] {
\small transform
};

\node (Finished) [stx, below of=Transform, node distance=.8cm] {
\codefontsize \stx{target lang stx}
};


\draw [arrow] (Start) -- (Typecheck);
\draw [arrow] (Typecheck) -- (Transform);
\draw [arrow] (Transform) -- (Finished);
\end{tikzpicture}
\end{center}
\end{minipage}
%
%
%
\begin{minipage}[c]{0.458\textwidth}
\begin{alltt}\codefontsize
(\defmac \defname{stx-transform-name}
 [\pat{<input: surface lang stx pattern>}
  <side-condition checks>
  \stx{<output: construct target lang stx>}])
\end{alltt}
\end{minipage}
%
%
%
\begin{minipage}[c]{0.28\textwidth}
\begin{center}
\tikzstyle{pat} = [rectangle, rounded corners, minimum width=1cm, minimum height=0.1cm,text justified]
\tikzstyle{stx} = [rectangle, rounded corners, minimum width=1cm, minimum height=0.1cm,text justified]
\tikzstyle{rect} = [rectangle, rounded corners, minimum width=3cm, minimum height=0.1cm,text justified, draw=black]
\tikzstyle{arrow} = [->,>=stealth, line width=1pt]
\tikzstyle{line} = [-,>=stealth, line width=1pt]

\begin{tikzpicture}[auto, node distance=2cm,>=latex']
\node (Start) [pat] {\small \pat{surface lang stx}};

\node(TypecheckTransform) [rect, below of=Start, node distance=1.1cm] {
\begin{tabular}{c}
\small macro expand:\\
\small type check $\rightleftharpoons$ transform
\end{tabular}
};

\node (Finished) [stx, below of=TypecheckTransform, node distance=1.1cm] {
\small \stx{target lang stx}
};
\draw [arrow] (Start) -- (TypecheckTransform);
\draw [arrow] (TypecheckTransform) -- (Finished);

\end{tikzpicture}
\end{center}
\end{minipage}
\figcapspacing
\caption{(l) A typical type system implementation, (c) a macro definition, and (r) a macro-based type checker.}
\label{fig:tcarch}
\figpostspacing
\end{figure}
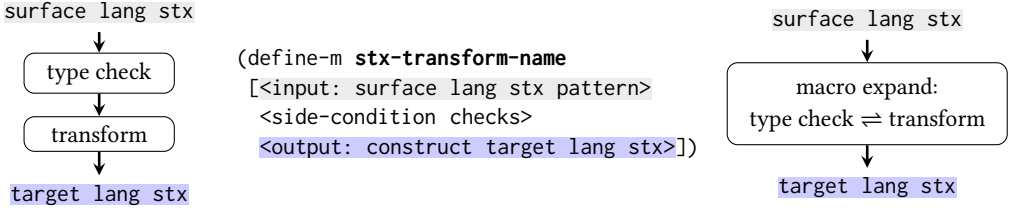

\fullref[]{fig:stlc} presents such rules for the simply typed
$\lambda$-calculus, split into bidirectional ``synthesize'' ($\synth$)---the
type is the output---and ``check'' ($\chck$)---the type is the
input---variants. More specifically a judgment $\Gamma\vdash
e\gg\overline{e}\Rightarrow\tau$ reads ``in context $\Gamma$, $e$ transforms to
$\overline{e}$ and has a type that matches pattern $\tau$, and $\Gamma\vdash
e\gg\overline{e}\Leftarrow\tau$ reads ``in context $\Gamma$, $e$ transforms to
$\overline{e}$ and has type equal to $\tau$. This kind of ``check and
transform'' rule, \eg, from~\citet{bidir-popl1998}, is a common way to
specify type systems.  In our~\fullref[]{fig:stlc} example, the ``transform''
part is a basic type erasure. Since tracking and manipulating the binders of a
program are important for type checking, \eg, when computing substitution or
\textalpha-equality, the rules conservatively generate fresh binders for the
target language to distinguish it from source language binders. This paper uses
an ``overline'' convention to distinguish source and target language
constructs. For example, a surface $\lambda$ and (an
explicitly-named) \textrm{app} function application transform to
$\overline{\lambda}$ and $\er{\textrm{app}}$, respectively, in some (for now
unspecified) target language. The next section shows how one might
implement~\fullref[]{fig:stlc}'s specification with macros.

%
%
%
\newcommand{\erasefn}{\mathit{erase}}
\begin{figure}[t]
\begin{mathpar}
\vspace{-10pt}\infer[App$\Rightarrow$]{\Gamma \vdash f \gg \overline{f} \Rightarrow \tau_1 \rightarrow \tau_2 \\ 
             \Gamma \vdash e \gg \overline{e}  \Leftarrow \tau_1}
            {\Gamma \vdash \textrm{app} f e \gg 
                           \overline{\textrm{app}}\, \overline{f}\, \overline{e}
               \Rightarrow \tau_2}

\infer[Lam$\Rightarrow$]{\Gamma, x \gg \overline{x} : \tau_1 \vdash e \gg \overline{e} \Rightarrow \tau_2
             \quad\overline{x}\not\in\textrm{\small dom}(\Gamma)}
            {\Gamma \vdash \lambda x:\tau_1.e \gg
                           \overline{\lambda} \overline{x}.\overline{e}
               \Rightarrow \tau_1 \rightarrow \tau_2}

\vspace{-8pt}\infer[App$\Leftarrow$]{\Gamma \vdash f \gg\overline{f} \Rightarrow \tau_1\rightarrow\tau_2
             \quad\tau_2=\tau_0\quad
             \Gamma \vdash e \gg\overline{e}  \Leftarrow \tau_1}
            {\Gamma \vdash \textrm{app} f e \gg 
                           \overline{\textrm{app}}\, \overline{f}\, \overline{e}
                \Leftarrow \tau_0}

\infer[Lam$\Leftarrow$]{\Gamma, x\gg\overline{x}:\tau_1\vdash e \gg\overline{e} \Leftarrow \tau_2
             \quad\overline{x}\not\in\textrm{\small dom}(\Gamma)}
            {\Gamma \vdash \lambda x.e \gg\overline{\lambda}\overline{x}.\overline{e}\Leftarrow \tau_1 \rightarrow \tau_2}

\infer[Var$\Rightarrow$]{x\gg\overline{x}:\tau\in\Gamma}
            {\Gamma \vdash x \gg\overline{x}\Rightarrow \tau}

\infer[Var$\Leftarrow$]{x\gg\overline{x}:\tau\in\Gamma\quad\tau=\tau_0}
            {\Gamma \vdash x \gg\overline{x} \Leftarrow \tau_0}
\end{mathpar}
%
\figcapspacing
\caption{Bidirectional ``check and transform'' rules for the simply typed $\lambda$-calculus.}
\label{fig:stlc}
\figpostspacing
\end{figure}

%
%
\subsection{From Specification To Implementation}


\fullref[]{fig:stlcimpl} presents two versions of a Racket \emph{module} named
\modu{stlc}, each containing an implementation of~\fullref[]{fig:stlc}'s
specifications. A \hlang{}at the top of a Racket module declares the language
of that module's code; thus,~\fullref[l]{fig:stlcimpl} depicts \modu{racket}
code. This \modu{racket} implementation of \modu{stlc} consists of two (type
checking) macro definitions, \apptt and \lm, each with two cases corresponding
to analogous $\synth$ and $\chck$ rules from~\fullref[]{fig:stlc}.\footnote{We
sometimes use square brackets, which are semantically equivalent to standard
parentheses, to help readability.}  The \apptt macro's first case implements
the $\textsc{App}\!\chck$ rule; thus its input must additionally include an
``expected'' type from its context\footnote{Like ``this'' from
OOP, \ttt{\scriptsize this-stx} is a macro's syntax object input. It may have ``expected'' type information from the context.} which binds it
to a \pat{\typ[0]} pattern variable.\footnote{The \texttt{\scriptsize\stxwhen}
and \texttt{\scriptsize\stxwith} directives specify additional side conditions
for a macro. Expansion may only proceed if predicates
following \texttt{\scriptsize\stxwhen} are true, and when the pattern
after \texttt{\scriptsize\stxwith} matches the syntax computed by the
subsequent expression.} The rest of the case follows straightforwardly from the
rest of $\textsc{App}\!\chck$. First, it computes the type of the applied
function \ttt{f}\footnote{Here we use an explicit quasiquote
constructor \texttt{\scriptsize\#\`}, which creates a syntax object according
to the template that follows it.}, and its erasure, with a call to a \syntt
function (\syntt and other helper functions are explained
in~\fullref[]{sec:primer:helperfns}). The
pattern \pat{(\arp\, \typ[1] \typ[2])} constrains the type of \ttt{f} to be a
function with one input and one output type (implementation of function types
is explained in~\fullref[]{sec:video}). If \ttt{f}'s type fails to match this
pattern, type checking fails with an error.  Next, the \apptt macro must check
that the output type of the function is equivalent to the expected type.  Then
the macro checks that the argument \ttt{e} has type \ttt{\ty[1]} using function
\chktt, which returns \ttt{\er{e}}, the erasure of \ttt{e}.  Finally, the macro emits an
erased term \ttt{(\er{\app} \er{f} \er{e})}.

%
%
%
%
\begin{figure}[t]
%
%
\begin{minipage}[t]{0.52\textwidth}
\begin{alltt}\codefontsize
\hlangrack   (provide \lm \app)  \moduname{stlc}
(\defmac \defname{\app}
 [\pat{(\app f e)} #:when (has-expected-\ty? this-stx)
  #:with \pat{\ty[0]} (get-expected-\ty this-stx)
  #:with \pat{[\erp{f} (\arp \typ[1] \typ[2])]} (\syntt #`\stx{f})
  #:when (\ty= #`\stx{\tyx[2]} #`\stx{\tyx[0]})
  #:with \pat{\erp{e}} (\chktt #`\stx{e} #`\stx{\tyx[1]})
  #`\stx{(\erx{\app} \erx{f} \erx{e})}]
 [\pat{(\app f e)}
  #:with \pat{[\erp{f} (\arp \typ[1] \typ[2])]} (\syntt #`\stx{f})
  #:with \pat{\erp{e}} (\chktt #`\stx{e} #`\stx{\tyx[1]})
  (\asstt #`\stx{(\erx{\app} \erx{f} \erx{e})} #`\stx{\tyx[2]})])
\vspace{4pt}(\defmac \defname{\lm}
 [\pat{(\lm x e)} #:when (has-expected-\ty? this-stx)
  #:with \pat{(\arp \typ[1] \typ[2])} (get-expected-\ty this-stx)
  #:with \pat{[\erp{x} \erp{e}]} (\chktt #`\stx{e} #`\stx{\ty[2]} #:ctx #`\stx{[x\::\:\tyx[1]]})
  #`\stx{(\erx{\lmx} (\er{x}) \erx{e})}]
 [\pat{(\lm [x : \typ[1]] e)}
  #:with \pat{[\erp{x} \erp{e} \typ[2]]} (\syntt #`\stx{e} #:ctx #`\stx{[x\::\:\tyx[1]]})
  (\asstt #`\stx{(\erx{\lmx} (\er{x}) \erx{e})} #`\stx{(\arx \tyx[1] \tyx[2])})])
\end{alltt}

\end{minipage}
%
%
\begin{minipage}[t]{0.47\textwidth}
\begin{alltt}\codefontsize
\hlang[\Turntwo](provide \lm \app)\moduname{stlc}
(\deftyrule \defname{\app}
 [\pat{(\app f e)} \chck \pat{\typ[0]} \expa
  \premss{f}{(\arp \typ[1] \typ[2])} [\stx{\tyx[2]} \ty= \stx{\tyx[0]}]
  \premcc{e}{\tyx[1]}
  -------------------------------
  \concc{(\erx{\app} \erx{f} \erx{e})}]
 [\pat{(\app f e)} \expa
  \premss{f}{(\arp \typ[1] \typ[2])}
  \premcc{e}{\tyx[1]}
  -------------------------------
  \concs{(\erx{\app} \erx{f} \erx{e})}{\tyx[2]}])
\vspace{4pt}(\deftyrule \defname{\lm}
 [\pat{(\lm x e)} \chck \pat{(\arp \typ[1] \typ[2])}\expa
  \premce[\bix{\tyx[1]}]{\tyx[2]}
  --------------------------------
  \concc{(\erx{\lm} (\erx{x}) \erx{e})}]
 [\pat{(\lm [x : \typ[1]] e)} \expa
  \premse[\bix{\tyx[1]}]{\typ[2]}
  --------------------------------
  \concs{(\erx{\lm} (\erx{x}) \erx{e})}{(\arx \tyx[1] \tyx[2])}])
\end{alltt}

\end{minipage}
%
\figcapspacing
\caption{STLC implementation, (l) using Racket, and (r) using \Turntwo.}
\label{fig:stlcimpl}
\figpostspacing
\end{figure}

Like $\textsc{App}\!\Rightarrow$ and $\textsc{App}\!\Leftarrow$, the second
\apptt macro case is similar to the first. The difference is that there is no
incoming ``expected'' type. Instead, the macro emits a second output, the type
of the function application term, which is the \ttt{\ty[2]} from the function
type.

In any syntax template, identifiers bound by in-scope syntax patterns refer to
the pieces of syntax matched in those patterns. For example, in \stx{(\er{\app}
  \er{f} \er{e})}, \stx{\er{f}} refers to \patdk{\er{f}} from a previous
pattern. Identifiers not bound in a previous pattern, however, reference
bindings from the rule's definition context. Thus \ttt{\er{\app}} in the \apptt
macro's output is \modu{racket}'s function application form. (We also
  use the overline convention to distinguish untyped Racket forms from any
  typed counterparts we may define.) The \ttt{\#\%} prefix naming convention
indicates an implicit form so programmers do not write \apptt explicitly;
instead, the macro expander automatically inserts it in front of applied
functions. So the \modu{stlc} module, by redefining \ttt{\app} (and exporting
it), \emph{changes} the behavior of function application.

\fullref[l]{fig:stlcimpl}'s {\lm} macro definition implements the \textsc{Lam}
rules from~\fullref[]{fig:stlc}, in much the same way as \apptt. The
interesting part of {\lm} is that it calls \chktt and \syntt with an extra
context argument consisting of the binding \ttt{x} and its type \ty. These
functions return one more result as well, the new ``erased'' binder, which is
used to construct the output term.

In the Racket ecosystem, \emph{a module is a language implementation}, and the
language's constructs are exactly the module's exports. Since \modu{stlc}
exports \ttt{\app} and {\lm}, writing \hlang[stlc] at the top of a module
means that the subsequent code is type checked by
\modu{stlc}'s macros. Obviously, programmers cannot write any \modu{stlc} code yet
because we have not defined any types; in general, however, this ability to
control language features---users of a particular \ttt{\#lang} cannot arbitrarily
access features from another one---is useful when implementing dependent types,
where unrestrained language interoperation can accidentally introduce
inconsistency in the underlying logic.

The close correspondence between~\fullref[]{fig:stlc}'s specification
and~\fullref[l]{fig:stlcimpl}'s \modu{racket} implementation suggests that
programmers could implement typed languages using syntax closer
to~\fullref[]{fig:stlc}. \fullref[r]{fig:stlcimpl} shows \modu{stlc},
implemented with \Turntwo. Though the two sides of~\fullref[]{fig:stlcimpl}
somewhat resemble each other---the same patterns and templates are just
rearranged---the extra layer is important for usability and reasoning while
programming. For example, the left uses explicit macro operations and thus
report errors in this low-level language, \eg, \textit{``bad syntax''}. In
contrast the right embeds implicit patterns and templates in a language of type
rules, and these abstractions enable more domain-appropriate
errors, \eg, \textit{``type mismatch, expected X, got Y''}. Since \Turntwo
sugars the calls to
\syntt and \chktt functions with the syntax of the judgments
from~\fullref[]{fig:stlc}, the rules are read similarly: the {\premset} and
{\premcet} ``premise judgments'' are read, respectively, ``\stx{e} transforms
to a term matching pattern \pat{\erp{e}} and has type matching pattern
\pat{\typ}'', and ``\stx{e} rewrites to a term matching pattern \pat{\erp{e}}
and has type \stx{\tyx}''. The ``conclusion judgment'' of a \ttt{\deftyrule},
is split into its input components at the top, corresponding to the macro's
input, and the output components at the bottom, corresponding to the macro's
outputs. This allows the rule implementation to be read like code, top to
bottom.

%
%
\subsection{Helper Functions}\label{sec:primer:helperfns}

Despite their similarities, there are key differences between
Figures~\ref{fig:stlc} and~\ref{fig:stlcimpl}(r): the latter has
no \textsc{Var} rule nor explicit $\Gamma$ environments. To understand these
discrepancies, we look at implementations of the \syntt, \chktt, and \asstt
functions, in~\fullref[]{fig:metafns}, which reveal how
\Turntwo reuses Racket's macro infrastructure to implement these missing parts,
and to more generally facilitate type checking. These helper functions require
only a few lower-level operations on syntax; thus our entire type checker is
implemented ``as macros''. Specifically, the functions rely on three macro
system features: (1) a programmatic way to add macro definitions to the macro
environment, (2) a \expatt function that manually initiates macro expansion on a syntax
object; and (3) a way of associating additional information (\eg, types) with syntax
objects; we use \emph{syntax properties} which, via \attname and
\detname, associate key-value pairs to syntax objects.

\begin{figure}[t]
\begin{minipage}[t]{0.47\textwidth}
\begin{alltt}\codefontsize
(define (\asstt e \ty) \att{e}{type}{\ty})
(define (\syntt e \stxctx \pat{[x : \typ]})
 (define \er{x} (fresh))
 (define env
  (envadd-m (curr-env) #`\stx{x} (\asstt \er{x} #`\stx{\ty})))
 (define \er{e} (local-expand e env))
 (define \ty[e] \dett{\er{e}}{type})
 #`\stx{(}#,\er{x}\stx{ }#,\er{e}\stx{ }#,\ty[e]\stx{)})
\end{alltt}
\end{minipage}
\begin{minipage}[t]{0.52\textwidth}
\begin{alltt}\codefontsize
\moduname{\Turntwo}
(define (has-expected-\ty? e) (has-prop? e \textrm{\textit{'{}exp}}))
(define (add-expected-\ty e \ty) \att{e}{exp}{\ty})
(define (get-expected-\ty e) \dett{e}{exp})
(define (\chktt e \ty \stxctx ctx)
  (define e/exp (add-expected-\ty e \ty))
  #:with \pat{[\erp{x} \erp{e} \typ[e]]} (\syntt e/exp \stxctx ctx)
  (if (\ty= #`\stx{\tyx[e]} \ty) #`\stx{(\erx{x} \erx{e})} (err "ty mismatch")))
\end{alltt}
\end{minipage}
\figcapspacing
\caption{``Type systems as macros'' core API.}
\label{fig:metafns}
\figpostspacing
\end{figure}

Individually, \asstt attaches a type to a term, at key \textit{'{}type}. The
``expected type'' functions use the same API at key \textit{'{}exp}. The \syntt
function consumes an expression \ttt{e} and an (optional) \ttt{ctx}---which
has shape \pat{[x : \typ]}---representing bindings to add to a type
environment. The \syntt function first generates a fresh binding \ttt{\er{x}}
(corresponding to \ttt{\er{x}} in figures~\ref{fig:stlc}
and~\ref{fig:stlcimpl}). Then it creates a new macro environment instance, which
extends the old one with a \emph{new macro definition} named
\ttt{x}. This new macro expands to \ttt{\er{x}} and its type \ty. In this way,
the \textsc{Var} rules from~\fullref[]{fig:stlc} are implemented in exactly the
same way as \ttt{\lm} and \ttt{\app}: as a macro that expands to the term and
type outputs of the ``type check and transform'' relation. This also explains
the lack of $\Gamma$s in~\fullref[]{fig:stlcimpl}; since our macro-based type
checking \emph{re-uses Racket's macro environment as the type environment},
scoping of type environment bindings is automatically handled by the macro
expander. A programmer need only specify the \emph{new} bindings like for
{\lm}.  The environment structure with these new bindings are then passed to
\expatt, which invokes the appropriate type checking macro for \ttt{e}.  After
expanding (\ie, type checking and transforming) \ttt{e}, its \prop{type} syntax
property is retrieved and \syntt returns a triple, as syntax, containing the
fresh name, the transformed \ttt{\er{e}}, and its type (the quasiquotation
\ttt{\#,} escape operator allows splicing metalanguage terms).

The \chktt function first invokes \syntt on term \ett, checks that the actual
and expected type match using type-equality function \tytt=, and returns the
expanded $\ertt$ if successful. For this paper, we assume \tytt= (not shown) is
syntactic equality up to \textalpha-equivalence; it is straightforward to implement
directly on the concrete syntax, \ie, we need not convert to an alternate
representation like deBruijn indices, since syntax objects are already aware of
the program's binding structure.

\fullref[]{fig:callgraph} shows a macro call graph for an \modu{stlc} example,
\ttt{({\lmtt} [x : Int] (add1 x))}, where we assume \fullref[]{fig:stlcimpl} is
extended with arrow and \ttt{Int} types, and an \ttt{add1} primitive with type
\ttt{({\artt} Int Int)}. Expansion (and type checking) begins with the
\ttt{\lmtt} macro, whose invocation calls the \ttt{\syntt} function, which in
turn calls \expatt with the lambda body, which invokes the next type checking
macro, \ttt{\app}. Return edges are marked with macro outputs, \ie, expanded
terms and their type (attached as a syntax property), so the example has final
type \ttt{({\artt} Int Int)}.

\newcommand{\pictt}[1]{\texttt{\scriptsize{}#1}}
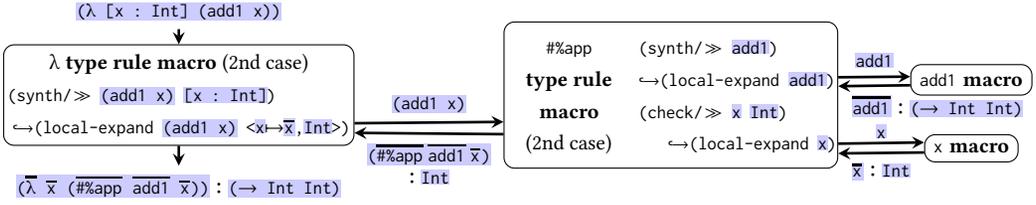
\begin{figure}[t]
\tikzstyle{pat} = [rectangle, rounded corners, minimum width=.9cm, minimum height=0.1cm,text justified]
\tikzstyle{stx} = [rectangle, rounded corners, minimum width=.9cm, minimum height=0.1cm,text justified]
\tikzstyle{rect} = [rectangle, rounded corners, minimum width=2cm, minimum height=0.1cm,text justified, draw=black]
\tikzstyle{rectsm} = [rectangle, rounded corners, minimum width=1cm, minimum height=0.1cm,text justified, draw=black]
\tikzstyle{arrow} = [->,>=stealth, line width=1pt]
\tikzstyle{arrow2} = [<-,>=stealth, line width=1pt]
\tikzstyle{line} = [-,>=stealth, line width=1pt]

\begin{tikzpicture}[auto, node distance=2cm,>=latex']
\node (Start) [pat] {
\pictt{\stxsm{({\lmtt} [x : Int] (add1 x))}}
};

\node(Lam) [rect, below of=Start, node distance=1.1cm,align=center] {
\footnotesize{\lmtt} \textbf{type rule macro} (2nd case) \\
\hspace{-28pt}\pictt{({synth/\ensuremath{\gg}} \stxsm{(add1 x)} \stxsm{[x : Int]})} \\
{\tiny$\hookrightarrow$}\pictt{(local-expand \stxsm{(add1 x)} <\stxsm{x}\ensuremath{\tiny\mapsto}\stxsm{\er{x}},\stxsm{Int}>)}\hspace{-2pt}
};

\node (Finished) [stx, below of=Lam, node distance=1.2cm] {
\pictt{\stxsm{(\er{\lmtt} \er{x} (\er{\app} \er{add1} \er{x}))}} : \pictt{\stxsm{(${\tiny\rightarrow}$ Int Int)}}
};

\node(App) [rect, right of=Lam, node distance=6.55cm, align=center] {
\begin{tabular}{cl}
{\footnotesize \pictt{{\app}}} &
{\pictt{({synth/\ensuremath{\gg}} \stxsm{add1})}} \\
\textbf{\footnotesize type rule} &
{\tiny$\hookrightarrow$}{\pictt{(local-expand \stxsm{add1})}\hspace{-8pt}} \\
\textbf{\footnotesize macro} &
{\pictt{({check/\ensuremath{\gg}} \stxsm{x} \stxsm{Int})}} \\
{\footnotesize (2nd case)} & 
\hspace{11pt}{\tiny$\hookrightarrow$}{\pictt{(local-expand \stxsm{x})}}\hspace{-8pt}
\end{tabular}};

\node(Y) [pat, right of=App, node distance=4cm] {
};

\node(Add) [rectsm, above of=Y, node distance=.2cm] {
\pictt{add1} \textbf{\footnotesize macro}
};

\node(X) [rectsm, below of=Y, node distance=.7cm] {
\pictt{x} \textbf{\footnotesize macro}
};

\draw [arrow] (Start) -- (Lam);
\draw [arrow] (Lam) -- (Finished);
\draw [arrow]  (Lam.351)   -- node[above]{\pictt{\stxsm{(add1 x)}}}         (App.189); 
\draw [arrow2] (Lam.348) -- node[below]{\pictt{\stxsm{(\er{\app}\:\er{add1}\:\er{x})}}}
                            node[below,yshift=-10pt]{: \stxsm{\pictt{Int}}} (App.193);
%
\draw [arrow]  (App.6) -- node[above]{\pictt{\stxsm{add1}}} (Add.177); 
\draw [arrow2] (App.3) -- node[below,xshift=24pt]{\pictt{\stxsm{\er{add1}}} : \pictt{\stxsm{(${\tiny\rightarrow}$ Int Int)}}} (Add.185);
%
\draw [arrow]  (App.344) -- node[above,yshift=-2pt]{\pictt{\stxsm{x}}} (X.173); 
\draw [arrow2] (App.341) -- node[below]{\pictt{\stxsm{\er{x}}} : \pictt{\stxsm{Int}}} (X.185); 
\end{tikzpicture}
\figcapspacing
\caption{Macro Call Graph For a Basic \modu{stlc} Example}
\label{fig:callgraph}
\figpostspacing
\end{figure}

%% file: video.tex
\citet{macrotypes}'s original \Turnstile could not handle dependent types since
it assumes that terms and types are distinct. We introduce how \Turntwo covers
this deficit with
\modu{Typed Video}, a DSL with \emph{indexed types}---``lightweight'' dependent
types in the style of Dependent ML~\cite{dependentml}---implemented ``as
macros''. With indexed types, we can lift some terms (the index language) to
the type level to express simple predicates about those terms. While
\citet{video} introduced \modu{Typed Video} and briefly describe a few type
rules, our work is the first to explain the underlying implementation details
of such rules and their accompanying types. (As \modu{Typed Video} is not our main
focus, we do ignore unrelated parts of the language, \eg, the details of constraint solving.)

\newcommand{\Prod}{Prod} 

\modu{Typed Video} is a typed version of~\citet{video}'s \modu{video} language,
a DSL for editing videos that has been used to create the video proceedings of
summer schools like OPLSS and conferences like POPL.
\modu{Typed Video}'s indexed types statically rule out errors that arise when
creating and combining video streams. More specifically, a \modu{Video} program
manipulates \emph{producers}---streams of data such as audio, video, or some
combination thereof---cutting, splicing, and mixing them together into a final
product. Since video editing is a multi-phase process, errors---\eg,
accidentally using more data than exists---often do not manifest until late in
the editing process, usually during rendering, making them hard to find. To
catch these problems earlier, \modu{Typed Video} assigns producer values
a \ttt{\Prod} type, indexed by its length.
%
%
This is an ideal type system since (even untyped) video editing already requires
annotating many expressions with their length.

Below is a function that combines audio, video, and slides to create a
conference talk video.
\begin{alltt}\codefontsize
\hlang[typed/video]\moduname{video-prog}
(\defdef (\defname{mk-conf-talk} [n : Nat] [aud : (\Prod n)] [vid : (\Prod n)] [slide : (\Prod n)])
                               \hfill   -> (\Prod (+ n 6)) #:when (> n 3)\hspace{10pt}
\end{alltt}
\vspace{-8pt}
\begin{alltt}\codefontsize
  (playlist (img "conf-logo.png" #:len 6)
            (fade #:len 3)
            (overlay aud vid slide)))
\end{alltt}
\newcommand{\arvidtt}{\ensuremath{\artt\sbtt{vid}}}
%
The \ttt{mk-conf-talk} function consumes an integer length \ttt{n} and audio,
video, and slide producers with types \ttt{({\Prod} n)}, meaning they must be
at least \ttt{n} frames long. The function combines its inputs with
\ttt{overlay}, and further adds a logo that fades into the main content. The
function specifies an additional constraint \ttt{({\textgreater} n 3)}, to
ensure that the inputs contain enough data to perform the fade
transition. Finally, the output type specifies that the input is extended by 6
frames, to account for the added logo. The function is assigned a binding
function type, where each argument is named:
\begin{alltt}\codefontsize
(\arvidtt [n : Nat] [a : (\Prod n)] [v : (\Prod n)] [sl : (\Prod n)] (\Prod (+ n 6)) #:when (> n 3))
\end{alltt}
The type of each argument may also reference the names preceding it, \ie, it
is a \emph{telescope}. The next subsection shows how one may implement such a
type with \Turntwo.

\subsection{A Core for \modu{Typed Video}, Now With Types}\label{sec:video:defty}

\fullref[]{fig:depvid} shows \modu{Typed Video}'s core, implemented with
\Turntwo; it resembles STLC from~\fullref[r]{fig:stlcimpl} except extra
\ttt{\defty} rules specify how to construct valid types. For example, the rule
for \ttt{Nat} says that a \ttt{Nat} instance has ``type''
\ttt{Type}.\footnote{We assume \ttt{\scriptsize{Type\,:\,Type}}
  here. \fullref[]{sec:precur}'s \Cur implements a hierarchy but this paper
  does not discuss these theoretical choices since they are largely orthogonal
  to our implementation techniques.} The \ttt{\Prod} rule for producers says
that constructing a \ttt{Prod} type requires a \ttt{Nat}, \ie, a \emph{a term},
argument. Finally, the \ttt{\arvidtt} specifies \emph{binders}, which
subsequent arguments may reference. The \ttt{\lm} rule (we only show ``synth''
cases) shows how to create terms with \ttt{\arvidtt} types. In the first case,
the parameter is a \ttt{Nat}, which may be referenced in both \ttt{e} \emph{and
  its type}. Only \ttt{Nat} terms may be lifted, however, so a second \ttt{\lm}
case handles non-\ttt{Nat} parameters: its output $\arvidtt$ is constructed
with a \ttt{fresh} dummy parameter (note the quasiquoting), which is not
referenced in \ttt{\er[2]{\ty}}. (The rule checks \ttt{\er[2]{\ty}} a
second time, to ensure that it does not reference \ttt{x}.) The \ttt{\app} rule
shows how \ttt{Nat} arguments are lifted to types: the conclusion replaces, in
\ttt{\er[2]{\ty}}, references to the \ttt{\er{x}} binder with the application
argument \emph{term}.\footnote{This single-case rule, used in the rest of
  paper for brevity, combines the rule name and input pattern. For rules like
  these without a ``check'' clause, \Turntwo uses a default ``check''
  implemented with ``synth'' and \ttt{\ty=}.}  Both rules transform into
untyped \modu{Video} terms.

\begin{figure}[t]
\begin{minipage}[t]{0.45\textwidth}
\begin{alltt}\codefontsize
\hlang[\Turntwo]
(\defty \defname{Nat} : Type)
(\defty \defname{Prod} : Nat -> Type)
(\defty \arbf\sbtt{\bf{\texttt{vid}}}
      \#:bind [X : Type] : Type -> Type)

(\deftyrule \pat{(\defname{\app} f \ett)} \expa
  \premss{f}{(\arvidtt [\er{x} : \ertytt[1]] \ertytt[2])}
  \premce{\ertytt[1]}
  ------------------------------------
  \concs{(\er{\app}\sbtt{vid} \er{f} \ertt)}{\colorbox{white}{(subst \stx{\er{e}} \stx{\er{x}} \stx{\er[2]{\ty}})}})
\end{alltt}
\end{minipage}
\begin{minipage}[t]{0.54\textwidth}
\begin{alltt}\codefontsize
\moduname{typed/video}
(\deftyrule \defname{\lm}
 [\pat{(\_ [x \lit{:} \tytt[1]] \ett)} \expa \comm{Nat case}
  \premc{\tytt[1]}{Nat}{\Tytt}
  \premss[\bix{Nat}]{\ett}{\er[2]{\ty}}
  -----------------------------
  \concs{(\er{\lm}\sbtt{vid} (\er{x}) \er{e})}{(\arvidtt [\er{x} : Nat] \er[2]{\ty})}]
 [\pat{(\_ [x \lit{:} \tytt[1]] \ett)} \expa
  \premcc{\tytt[1]}{\Tytt}
  \premss[\bix{\er[1]{\ty}}]{e}{\er[2]{\ty}} \premc{\er[2]{\ty}}{\_}{\Tytt}
  -----------------------------
  \concs{(\er{\lm}\sbtt{vid} (\er{x}) \er{e})}{(\arvidtt [\colorbox{white}{(fresh)} : \er[1]{\ty}] \er[2]{\ty})}])
\end{alltt}
\end{minipage}
\figcapspacing
\caption{\modu{Typed Video} type definitions, lambda, and function application rules.}
\label{fig:depvid}
\figpostspacing
\end{figure}

%
%
%
%
\subsection{Defining ``Type'' Rules For Types}\label{subsec:typedefs}

Sections~\ref{subsec:typedefs} to~\ref{subsec:patexpanders} explain how
\ttt{\defty} is implemented. A \ttt{\defty} definition specifies a checking
rule, for a type. We can implement such a rule with the previously seen \ttt{\deftyrule} because
\ttt{\deftyrule}'s checking of syntax does not actually know the difference between
``terms'' or ``types''. \fullref[]{fig:typedefs} shows what these explicit
\ttt{\deftyrule}s might look like for both plain and \modu{Typed Video}'s
function types.
The rule for a standard $\artt$ checks that its input and output have type
$\Tytt$. But what should be the ``transform'' output of the rule?  Typed {\lm}
transforms to a target language's $\er{\lm}$, but we have more freedom in
choosing a type's underlying representation. Considering desired operations on
types, however, reveals some criteria for such a representation. Specifically,
type checking requires computing binding-aware operations like
\textalpha-equality and capture-avoiding substitution. Since syntax objects know
a program's binding structure, using a binding-valid representation lets us
exploit the macro system to implement such operations for free. Thus our
criteria for a type representation: (1) uniquely identifies the type; (2) includes the
arguments to the type constructor; and 3) respects hygiene, \ie, \emph{it has a
  valid binding structure in the target language}. For the first two criteria,
we define a (named) record $\er{\artt}$, declared with \ttt{\defstruct}, and
represent $\artt$ with a \emph{syntax object} that applies $\er{\artt}$ to the
function type constructor's arguments, as seen
in~\fullref[l]{fig:typedefs}. For binding types like $\arvidtt$
in~\fullref[r]{fig:typedefs}, to satisfy the hygiene criteria, its
representation in the conclusion includes a $\er{\lm}$ that wraps and binds
references to \ttt{\er{x}} in \ttt{\er[2]{\ty}}.

\subsection{Type Checking Telescopes: A New \Turntwo Premise} \label{subsec:telescopecheck}

\begin{figure}[t]
\begin{minipage}{0.49\textwidth}
\begin{alltt}\codefontsize
(\defstruct \er{\arbf} (in out))
(\deftyrule \pat{(\defname{\arbf} \typ[1] \typ[2])} \expa
  \premctt[1]
  \premctt[2]
  ----------
  \concsT{(\erx{\app} \erx{\arx} \er[1]{\ty} \er[2]{\ty})})
\end{alltt}
\end{minipage}
\begin{minipage}{0.49\textwidth}
\begin{alltt}\codefontsize
(\defstruct \er[vid]{\arbf} (in out))
(\deftyrule \pat{(\defname{\arbf\sbtt{vid}} [x : \tytt[1]] \tytt[2])} \expa
 \premc{\tytt[1]}{\ertytt[1]}{\Tytt}
 \premc[\bix{\ertytt[1]}]{\tytt[2]}{\ertytt[2]}{\Tytt}
 -----------------------------
 \concs{(\er{\app} \er[vid]{\artt} \er[1]{\ty} (\er{\lm} (\er{x}) \ertytt[2]))}{\Tytt})
\end{alltt}
\end{minipage}
\figcapspacing
\caption{Explicit \ttt{\deftyrule} type rules for single-arity function types, (l) $\artt$, (r) $\artt\sbtt{vid}$}
\label{fig:typedefs}
\figpostspacing
\end{figure}
%

%
%
%
\begin{figure}[t]
\begin{minipage}[t]{0.42\textwidth}
\begin{alltt}\codefont
(\defstruct \er{\arbf} (ins out))
(\deftyrule \pat{(\arbf \typ[1] \ooo \typ[2])} \expa
  \premctt[1] \ooo
  \premctt[2]
  ----------
  \concsT{(\erx{\app} \erx{\arx} (\er[1]{\ty} \ooo) \er[2]{\ty})})
\end{alltt}
\end{minipage}
\begin{minipage}[t]{0.573\textwidth}
\begin{alltt}\codefont
(\defstruct \defname{\er{\arbf}\sbtt{vid}} (types))
(\deftyrule \pat{(\defname{\arbf\sbtt{wrong}} [x : \typ[1]] \ooo \typ[2])} \expa
 [\bix{\er[1]{\ty}}\ooo \turn \chkjT{\stx{\ty[1]}}{\pat{\er[1]{\ty}}} \ooo]
 [\bix{\er[1]{\ty}}\ooo \turn \chkjT{\stx{\ty[2]}}{\pat{\er[2]{\ty}}}]
 -----------------------------
 \concs{(\erx{\app} \er[vid]{\arx} (\lmx (\erx{x} \ooo) \er[1]{\ty} \ooo \er[2]{\ty}))}{\Tytt})
\defspacing(\deftyrule \pat{(\defname{\arbf\sbtt{vid}} [x : \typ[1]] \ooo \typ[2])}
  \premc[\depbic{x}{\tytt[1]}{\Tytt}\ooo ]{\tyx[2]}{\er[2]{\ty}}{\Tytt}
  -----------------------------
  \concs{(\erx{\app} \er[vid]{\artt} (\lmx (\erx{x} \ooo) \er[1]{\ty} \ooo \er[2]{\ty}))}{\Tytt})
\end{alltt}
\end{minipage}
\figcapspacing
\caption{Explicit \ttt{\deftyrule} type rules for multi-arity function types, (l) $\artt$, (r) $\artt\sbtt{vid}$}
\label{fig:typedefs2}
\figpostspacing
\end{figure}

Suppose we want multi-arity function types. The \ttt{\artt} rule
in~\fullref[l]{fig:typedefs2} uses an ellipsis pattern \patdk{\ooo}, which
means ``match the preceding pattern zero or more times'', to specify multiple
arguments. Any pattern variables in this preceding pattern, like
\patdk{\ty[1]}, must be accompanied with another ellipsis when used in a syntax
template, \eg, \stx{\ty[1]\ooo}. \Turntwo allows writing an ellipsis after a
premise, which automatically inserts corresponding ellipses for \emph{all}
patterns and templates in that premise. For example the first premise
in~\fullref[l]{fig:typedefs2} checks that each type in \stx{\ty[1]\ooo} has
type \ttt{Type}, and matches the expansion of those types to
\pat{\er[1]{\ty}\ooo}.

The $\artt\sbtt{wrong}$ rule in~\fullref[r]{fig:typedefs2} tries to use the
same ellipsis pattern as the left, but this is the wrong binding structure
because each \ttt{\ty[1]} is checked in a context with \emph{every} \ttt{x},
\emph{including its own}. It's wrong because the \ttt{\ooo} usually means
``map''.  But to type check the telescoping binding structure on the right, we
need a ``fold'' operation that \emph{interleaves binding with checking}. This
means we need a new kind of premise, seen in the multi-arity
\ttt{{\artt}\sbtt{vid}} rule in~\fullref[r]{fig:typedefs2}, which is a
corrected version of $\artt\sbtt{wrong}$. Specifically, a premise
$\depbic{x}{\tytt[1]}{\Tytt}\ooo$ checks each $\stx{\tytt[1]}$, but also names
it so that subsequent type checking invoked by the ellipsis may reference the
argument.

\newcommand{\stxeq}{stx\textalpha=}

This new premise requires revising~\fullref[]{fig:metafns}'s API to accommodate
the fold operation; the new API is in~\fullref[]{fig:foldingcheck}. The main
function is now \ttt{expand/bind}, which consumes an identifier \ttt{x}, a
syntax object \ttt{stx}, a \ttt{tag} at which to associate \ttt{x}
with \ttt{stx} (e.g., \prop{type}), and a (macro) environment \ttt{env}. This
new API is agnostic: we do not care whether \ttt{stx} is a type, term, or
something else. The \ttt{expand/bind} function expands \ttt{stx} in the context
of \ttt{env}'s bindings, and then adds \ttt{x} to \ttt{env} as a ``type rule''
macro where \ttt{x} expands to a fresh \ttt{\er{x}}, with \ttt{stx} attached at
some \ttt{tag}.  For rules that do not want to bind after checking,
\ttt{expand/bind} can be called with a dummy \ttt{x} and \ttt{tag}.
Folding ``check'' premises, however, would use \ttt{expand/bind/check}, which wraps \ttt{expand/bind}. It consumes
an additional argument \ttt{stxval}, and ``checks'' that expanding \ttt{stx}
results in a syntax object that has \ttt{tag} ``equal'' to
\ttt{stxval}. Concretely, the premise of~\fullref[r]{fig:typedefs2}'s \ttt{\arvidtt}
rule would fold \ttt{expand/bind/check} over \stx{x\ooo}, \stx{\tyx[1]\ooo},
and \stx{\Tytt\ooo}, where each \stx{\tyx[1]} is the \ttt{stx} argument
and \stx{\Tytt} is \ttt{stxval}. The fold would accumulate the \ttt{env}
argument which holds the resulting \pat{\er{x}\ooo}
and \pat{\er{\typ[1]}\ooo}. Where~\fullref[]{fig:metafns} used
\ttt{\ty=},~\fullref[]{fig:foldingcheck} uses $\boxed{\ttt{\stxeq}}$, which is
an agnostic equality function. The box denotes an \emph{interposition point};
\Turntwo implements several of these overloadable hooks at key points to enable
more extensible rules.

%
\begin{figure}[t]
\begin{minipage}[t]{0.415\textwidth}
\begin{alltt}\codefontsize
(\defdef (\defname{expand/bind} x stx tag env)
  (define \er{stx} (\expatt stx env))
  (env-add env x tag \er{stx}))
(\defdef (\defname{env-add} env x tag \er{stx})
  (define \er{x} (fresh))
  (env-add-m env x (attach \er{x} tag \er{stx})))
\end{alltt}
\end{minipage}
\begin{minipage}[t]{0.575\textwidth}
\begin{alltt}\codefontsize
(\defdef (\defname{expand/bind/check} x stx tag stxval env)
  (define env\sbtt{new} (expand/bind x stx tag env))
  (if (\boxed{\texttt{\stxeq}} (detach (lookup env\sbtt{new} x) tag) stxval)
      env\sbtt{new}
      (err "ty mismatch")))
\end{alltt}
\end{minipage}
\figcapspacing
\caption{Type- and term-agnostic, revised version of the API from \hyperref[fig:metafns]{Figure \ref{fig:metafns}}; supports interleaved bind and check.}
\label{fig:foldingcheck}
\figpostspacing
\end{figure}
%


\subsection{Putting it All Together}\label{subsec:patexpanders}

We need one more component for \ttt{\defty}. The rule outputs
in~\fullref[]{fig:typedefs2} are still somewhat arbitrary; a programmer should
not need to know this underlying representation. Instead, they should use a
type's \emph{surface} syntax, as~\fullref[]{fig:stlcimpl} does when pattern
matching with \patdk{\artt}. To enable this, we define ``pattern macros'' for
each type, which are used exclusively in syntax pattern positions:
%
\begin{alltt}\codefontsize
    (define-pattern-m \pat{(\defname{\arvidtt} [x : \tytt[1]] \ooo \tytt[2])} \stx{(\er{\app} \er{\artt}\sbtt{vid} (\er{\lm} (x \ooo) \tytt[1] \ooo \tytt[2]))})
\end{alltt}
%
%
This macro matches on, but hides, a type's internal representation. A pattern
macro may have the same name as the type because its expansion occurs
\emph{before}, \ie, in a different namespace from, a \ttt{\deftyrule}.  This
feature relieves \Turntwo programmers from a heavy notational burden.  With
pattern macros, along with a \ttt{\defstruct} declaration for the internal
representation, and a \ttt{\deftyrule} implementing the type constructor rule we
can implement \ttt{\defty}, which is just sugar for this collection of
definitions.

\subsection{Type-Level Computation}

Since \modu{Typed Video} types may contain expressions, \textalpha-equality
alone no longer suffices. Thus, we also define a type normalization
function, which is straightforward using the macro system's facilities
for manipulating syntax. To install this normalization function, we do not need
to change any rules; instead, we take advantage of another \Turntwo
interposition point, in \ttt{expand/bind}:
%
\begin{alltt}\codefontsize
    (\defdef (\defname{expand/bind} x stx tag env) (env-add env x tag (\boxed{\ttt{norm}} stx env)))
\end{alltt}
%
This new version modifies~\fullref[]{fig:foldingcheck}'s version by replacing
\ttt{\expatt} with an overloadable \param{norm} (that itself defaults to
\ttt{\expatt}). This ensures that only normal forms are used when computing
equality with \param{stx\textalpha=} in~\fullref[]{fig:foldingcheck}. For
\modu{Typed Video}, we implement an interpreter for the index language.
\begin{figure}[t]
\begin{alltt}\codefontsize
\hlang[\Turntwo]\moduname{typed/video}
(define-\boxed{\ttt{norm}}
 [\pat{n} #:when (nat-lit? \stx{n}) \stx{n}] \hspace{56pt}[\pat{b} #:when (bool-lit? \stx{b}) \stx{b}]
 [\pat{(+ n m)}    #:with \pat{\er{n}} (\boxed{\ttt{norm}} \stx{n})    #:with \pat{\er{m}} (\boxed{\ttt{norm}} \stx{m})
  (if (and (nat-lit? \stx{\er{n}}) (nat-lit? \stx{\er{m}})) (+ (stx->lit \stx{\er{n}}) (stx->lit \stx{\er{m}})) \stx{(+ \er{n} \er{m})})]
 [\pat{(Producer n)} \stx{(Producer }(\boxed{\ttt{norm}} \stx{n})\stx{)}] \quad[\pat{other} \stx{other}])
\end{alltt}
%
\figcapspacing
\caption{Excerpt of type-level evaluation in the \modu{Typed Video} language.}
\label{fig:videolang}
\figpostspacing
\end{figure}
\fullref[]{fig:videolang} conveys the basic idea. It uses
$\ttt{define-\boxed{\ttt{norm}}}$ to overload \param{norm}, which is implemented as a
series of pattern-body cases. The first
two cases match literal values. The third case matches on addition, recursively
calling \ttt{norm} on the arguments. If evaluating those terms produce
syntactic literal numbers, then the actual arithmetic operation is performed;
otherwise, normalization produces a normalized addition syntax object. This
fourth case is similar: if the input is a \ttt{Producer}, then its index is
normalized. Finally, the last case leaves the type unchanged.

%% file: dep.tex
The approach to type-level computation for \fullref[]{sec:video}'s \modu{Typed
  Video} suffices for a simple index language but this approach is not
  extensible, \eg, it breaks down for languages where introducing new datatypes
  is possible. This section presents a more general and extensible approach to
  adding type-level computation, which we dub \emph{normalization by macro
  expansion} because each reduction rule is implemented as a separate macro.
  We explain using a calculus with full-spectrum dependent types, comparable to
  the Calculus of Constructions (CC)~\cite{coquand1988}, in which there is no
  distinction between terms and types. We then extend this initial
  implementation with type schemas, \`a la Martin-L\"of Type
  Theory~\cite{martin-loef1975}, and finally add inductive types, demonstrating
  that our approach scales to the calculi used in contemporary proof
  assistants. Further, each extension is modular: it only defines new
  constructs and does not modify prior code.

\newcommand{\extrasp}{\vspace{-6pt}}
\extrasp\subsection{Defining Type-Level Reductions}
\begin{figure}[t]
\begin{minipage}{0.54\textwidth}
\begin{alltt}\codefontsize
\hlang[\Turntwo]
(\defty \defname{\Pitt} #:bind [X : Type] : Type -> Type)
(\deftyrule \pat{(\defname{\app} f \ett)} \expa
  \premss{f}{(\Pitt [\er{X} : \er[1]{\ty}] \er[2]{\ty})}
  \premce{\er[1]{\ty}}
  ------------------------------------
  [\turn \stx{(\betatt \er{f} \er{e})} \synth (\refl (subst \stx{\er{e}} \stx{\er{X}} \stx{\er[2]{\ty}}))])
\end{alltt}
\end{minipage}
%
\begin{minipage}{0.45\textwidth}
\begin{alltt}\codefontsize
\moduname{dep-lang}

(\deftyrule \pat{(\defname{\lm} [x : \tytt] \ett)} \expa
  [\turn \stx{\ty} \expa \pat{\er{\ty}} \chck \stx{\Tytt}]
  \premse[\bix{\er{\ty}}]{\er[2]{\ty}}
  ------------------------------
  \concs{(\er{\lm} (\er{x}) \er{e})}{(\Pitt [\er{x} : \er{\ty}] \er[2]{\ty})})
\end{alltt}
\end{minipage}
\begin{alltt}\codefontsize
(\defred \defname{\betar} \pat{(\er{\app} (\er{\lm} (x) body) e)} ~> (subst \stx{e} \stx{x} \stx{body}))
\end{alltt}
\figcapspacing
\caption{A full-spectrum dependently typed lambda calculus.}
\label{fig:dep}
\figpostspacing\extrasp
\end{figure}

This section introduces a \Turntwo construct called \ttt{\defred}, for defining
type-level computation. \fullref[]{fig:dep} presents \modu{dep-lang}, which
uses \ttt{\defred} to define a \ttt{\betar} reduction rule by specifying a
redex (as a syntax pattern) and a contractum (as a template). \modu{dep-lang}
also upgrades~\fullref[]{fig:stlcimpl}'s STLC by: (1) changing $\rightarrow$ to
{\textPi}, the dependent function type whose output type can refer to its
input; and (2) modifying {\lm} and \ttt{\app} to introduce and eliminate
{\textPi}. Since the \ttt{\betar} macro is invoked in \ttt{\app}'s output, type
computation is interleaved with type checking.

A \ttt{\defred} definition internally defines a macro that rewrites redexes into
contractums. \fullref[]{fig:beta} sketches what a \ttt{\betar} macro might look
like. If the first term is a \er{\lm} (first case), occurrences of parameter
\ttt{\er{x}} in the body are replaced with argument \ttt{\er{e}}. This
reduction may create more redexes in the contractum, \eg, if the \ttt{\er{e}}
argument is a function, so \ttt{\betar}'s first case applies
\ttt{\refl\sbtt{v1}}, which ``reflects'' \ttt{\er{\app}} references back to
\ttt{\betar}, enabling further reductions. Otherwise (second case), the result
of \ttt{\betar} is an unreduced \ttt{\er{\app}} \emph{neutral term}. This
\ttt{\betar} conceptually captures ``normalization by macro expansion'', but
it's still \emph{not extensible} since \ttt{\refl\sbtt{v1}} would need to know
about all possible reduction rules in advance.

Instead,~\fullref[]{fig:define-red} defines a new \ttt{\refl},
extensible via syntax properties. Instead of directly replacing
\ttt{\er{\app}}, \ttt{\refl} traverses a piece of syntax and checks for a
\textit{'reflect-name} property. If it exists, its value is used as
the reflected name. Correspondingly, \ttt{mk-reflected} creates such an
annotated syntax, for use in unreducible neutral terms, by attaching a
\textit{'reflect-name} property to a given \emph{placeholder} identifier. With
our function application example, the \ttt{placeholder} is \ttt{\er{\app}} and
the \ttt{reflid} is {\betar}.

\begin{figure}[t]
\begin{minipage}[c]{0.63\textwidth}
\begin{alltt}\codefontsize
(\defmac \defname{\betar}
  [\pat{((\er{\lm} (\er{x}) \er{body}) \er{e})} (\refl\sbtt{v1} (subst \stx{\er{e}} \stx{\er{x}} \stx{\er{body}}))]
  [\pat{(\er{f} \er{e})} \stx{(\er{\app} \er{f} \er{e})}])) \comm{neutral term}
\end{alltt}
\end{minipage}
\begin{minipage}[c]{0.36\textwidth}
\begin{alltt}\codefontsize
\comm{\ttt{\refl\sbtt{v1}}: fn mapping \ttt{\app} back to \ttt{\betar},}
\comm{NOT extensible}
(\defdef (\reflbf\sbtt{\bf{\texttt{v1}}} e) (subst \stx{\betatt} \stx{\er{\app}} e))
\end{alltt}
\end{minipage}
\figcapspacing
\caption{Possible \ttt{\betar} reduction rule, implemented as a plain macro, but \emph{not} extensible.}
\label{fig:beta}
\figpostspacing\extrasp
\end{figure}

\begin{figure}[t]
\begin{minipage}[t]{0.45\textwidth}
\begin{alltt}\codefontsize
(\defdef \reflbf \comm{extensible reflect fn}
 [\pat{x} #:when (and (id? \stx{x})(has-reflid? \stx{x}))
  \dett{\stx{x}}{reflect-name}]
 [\pat{(e \ooo)} \stx{((\refl e) \ooo)}]
 [\pat{otherwise} \stx{otherwise}])
(\defdef (\defname{mk-reflected} placehold reflid)
  \att{placehold}{reflect-name}{reflid})
\end{alltt}
\end{minipage}
\begin{minipage}[t]{0.54\textwidth}
\begin{alltt}\codefontsize
\comm{examples:}\moduname{\Turntwo}
\comm{\ttt{((mk-reflected \er{\app} \betatt) (\lm x x) (\lm x x))}}
\comm{\ttt{  = (\er{\app} (\lm x x) (\lm x x))}}
\comm{\ttt{(\refl ((mk-reflected \er{\app} \betar) (\lm x x) (\lm x x)))}}
\comm{\ttt{  = (\betar (\lm x x) (\lm x x))}}
\comm{\ttt{(\expatt }}
\comm{\ttt{ (\refl ((mk-reflected \er{\app} \betar) (\lm x x) (\lm x x))))}}
\comm{\ttt{  = (\lm x x)}}
\end{alltt}
\end{minipage}
%


\begin{alltt}\codefontsize
(\defmac \defname{\defred} \comm{\Turntwo form for defining reduction rules}
 [\pat{(\defred red-name redex ~> contractum)} \stx{(\defred red-name [redex ~> contractum])}] 
 [\pat{(\defred red-name [(placeholder redex-hd rst \ooo) ~> contractum] \ooo)} \comm{multi-redex case}
   \ultt(\defmac \defname{red-name}                                      \urtt
      [\pat{(redex-hd rst \ooo)} (\refl \stx{contractum})] \ooo
   \lltt  [\pat{(e \ooo)} \stx{(}(mk-reflected \stx{placeholder} \stx{red-name}) \stx{e \ooo)}]))\lrtt])
\end{alltt}
%
\figcapspacing
\caption{Extensible \Turntwo API for defining reduction rules, used to define \ttt{\betar}.}
\label{fig:define-red}
\figpostspacing\extrasp
\end{figure}


Finally, \ttt{\defred} in~\fullref[bot]{fig:define-red} is a macro-defining
macro that, given a redex pattern and contractum template, generates a macro
with the necessary calls to \ttt{\refl} and \ttt{mk-reflected}. Thus multiple
\ttt{\defred} declarations automatically cooperate with each other without
knowing of each other's presence.  The first case of \ttt{\defred} is shorthand
for single-redex reduction rules. It recursively calls the multi-redex second
case,\footnote{Since \texttt{\scriptsize{}\defred} is a macro-defining macro,
  it has nested syntax templates and patterns, making our usual coloring more
  difficult to see. To help, the outer syntax template in the second case is
  boxed with $\ulltt$ $\ulrtt$ instead of the usual blue text color.}  where
the generated macro definition, \ttt{red-name}, is a generalized version of
{\betar} from \fullref[]{fig:beta}. If this macro's inputs match the supplied
redex pattern, it rewrites them to the specified contractum, letting Racket's
macro patterns and templates automatically do the work. Otherwise, the result
is an unreduced neutral term with the supplied placeholder at the head, marked
with a \textit{'{}reflect-name} property. Thus if later reductions transform
the neutral term into a redex, \ttt{\refl} ensures that \ttt{red-name} is
invoked again to reduce it.

\extrasp\subsection{A Little Sugar}

\fullref[]{fig:dep}'s \modu{dep-lang} is roughly the Calculus of Constructions,
which is not a real programming language yet. Fortunately, languages created
with our approach are extensible via macros for free. We demonstrate this with
first some small sugar extensions, which we then use to create larger
extensions like type schemas and inductive datatypes.
%
\fullref[]{fig:sugar} defines currying, multi-argument forms \ttt{\textPi/c},
\ttt{\lm/c}, and \ttt{app/c}, which unroll into their univariate versions.
\ttt{\textPi/c} also allows omitting the binder when there is no dependency; in
this (third) case, a fresh identifier is used. This allows programmers to more
concisely use \ttt{\textPi} like the simply typed \ttt{\artt} when appropriate;
correspondingly it is doubly exported with this alternate \ttt{\artt} name. All
these macros use a ``dot'' pattern, which
matches a syntax object as a cons pair, split into its head and tail, \eg, in
\pat{(\textPi\;b\;.\;rst)}, \pat{b} is the first binder and \pat{rst} is the
rest of the type, which may include more binders. The \modu{dep-lang/sugar}
library exports these currying forms without their \ttt{/c} suffix, meaning
users of the library will have the original constructs overloaded with these
respective sugary forms.


\begin{figure}[t]
\begin{alltt}\codefont
\hlang[dep-lang]\moduname{dep-lang/sugar}
(provide/rename [\Pitt/c \Pitt] [\Pitt/c \artt] [\lm/c \lm] [app/c \app])
(\defmac \defname{\Pitt/c}  \hspace{2pt}[\pat{(\_ e)} \stx{e}] [\pat{(\_ [x : \ty] . rst)} \stx{(\Pitt [x : \ty] (\Pitt/c . rst))}]
                          [\pat{(\_ \ty . rst)}       \stx{(\Pitt [}(fresh)\stx{ : \ty] (\Pitt/c . rst))}])
(\defmac \defname{\lm/c}  \hspace{4pt}[\pat{(\_ e)} \stx{e}] [\pat{(\_ x+\ty . rst)} \stx{(\lmtt \hspace{2pt}x+\ty (\lmtt/c . rst))}])
(\defmac \defname{app/c} [\pat{(\_ e)} \stx{e}] [\pat{(\_ f e . rst)} \stx{(app/c (\app f e) . rst)}])
\end{alltt}
%
\figcapspacing
\caption{A \modu{dep-lang} library that adds some syntactic sugar for currying.}
\label{fig:sugar}
\figpostspacing\extrasp
\end{figure}

\extrasp\subsection{An Extension of Natural Numbers}

\begin{figure}[t]
\begin{alltt}\codefont
\hlang[\Turntwo]  (require \modu{dep-lang})  (provide Nat Z S \elim{Nat} \datum)  \moduname{dep-lang/nat}
\vspace{2pt}(\defty \defname{Nat} : Type) \quad(\defty \defname{Z} : Nat) \quad(\defty \defname{S} : Nat -> Nat)
\vspace{2pt}(\deftyrule \pat{(\defname{\elim{Nat}} n P mz ms)} \expa
  \premcc{n}{Nat} \comm{target}
  \premcc{P}{(\artt Nat Type)} \comm{prop / motive}
  \premcc{mz}{(\er{P} Z)} \comm{method for Z}
  \premcc{ms}{(\Pitt [k : Nat] (\artt (\er{P} k) (\er{P} (S k))))} \comm{method for S}
  -----------
  \concs{(\match{Nat} \er{n} \er{P} \er{mz} \er{ms})}{(\er{P} \er{n})})
\vspace{2pt}(\defred \defname{\match{Nat}}
  [\pat{(\match{Nat} Z P mz ms)} ~> \stx{mz}]
  [\pat{(\match{Nat} (S k) P mz ms)} ~> \stx{(ms k (\match{Nat} k P mz ms))}])
\vspace{2pt}(\defmac \defname{\datum}
  [\pat{(\_ n)} #:when (zero? \stx{n}) \stx{Z}]
  [\pat{(\_ n)} #:when (nat-lit? \stx{n}) \stx{(S (\datum\colorbox{white}{ (sub1 (stx->lit \stx{n}))}))}]
  [\pat{(\_ x)} \stx{(\er{\datum} x)}]))
\end{alltt}
%
\figcapspacing
\caption{A \modu{dep-lang} extension for natural numbers.}
\label{fig:nat}
\figpostspacing\extrasp
\end{figure}

To write interesting programs, \modu{dep-lang} needs more data types. Thus we
extend it---using already described tools and techniques---with a
MLTT-style~\cite{mlttbook} natural number type schema in \fullref[]{fig:nat}.
Like all \Turntwo constructs, \ttt{\defty} is not limited to defining ``types''
in the simply typed sense; instead, it's suitable for defining any
``constructor'' form. Thus this module uses it to define \ttt{Nat} \emph{and}
its introduction rules \ttt{Z} and \ttt{S}, corresponding to ``zero'' and
``successor''. The elimination form, \ttt{\elim{Nat}}, corresponds to a fold
over the datatype. Following the terminology of \citet{mcbride2000:thesis}, the
form \ttt{(\elim{Nat} n P mz ms)} takes \emph{target} to eliminate \ttt{n}, a
\emph{motive} \ttt{P} that describes the return type of this form, and one
\emph{method} for each case of natural numbers: \ttt{mz} when \ttt{n} is zero
and \ttt{ms} when \ttt{n} is a successor. Method \ttt{mz} must have type
\ttt{(P Z)}, \ie, the motive applied to zero, while \ttt{ms} must have type
$\ttt{(\Pitt\, [k : Nat] (\artt\, (P k) (P (S k))))}$, which mirrors an
induction proof: for any \ttt{k}, given a proof of \ttt{(P k)}, we show \ttt{(P
  (S k))}.  Using \ttt{\defred}, we can define reduction rules for \elim{Nat},
one each for \ttt{Z} and \ttt{S}. Observe that the pattern macros \patdk{Z} and
\patdk{S} (defined as part of \ttt{\defty}, explained
in~\fullref[]{subsec:patexpanders}) help specify the reduction succinctly.

Finally, the \ttt{Nat} module overloads the meaning of \emph{literal data} by
extending \ttt{\datum}, another Racket interposition point that wraps
literals. With the new \ttt{\datum}, users of the \modu{dep-lang/nat} can write
numeric literals in place of the more cumbersome \ttt{Z} and \ttt{S}
constructors. The last \ttt{\datum} clause falls back to a core $\er{\datum}$,
making this extension compatible with others that might extend literal data. We
could even support diamond extensions by importing \emph{two} existing versions
of $\datum$ (under different names) and use them in separate clauses of a new
\ttt{\datum}.

Notably, we use \ttt{\hlang[\Turntwo]} to implement this \ttt{Nat} module, not
\modu{dep-lang}, because (unlike~\fullref[]{fig:sugar}) this extension is
\emph{unsafe}. Ideally, the \modu{dep-lang} trusted core should not have the ability to add
new types and rules that could change the
logic. \fullref[]{sec:dep:trustedcore} discusses managing extensions.

%
%
%
\extrasp\subsection{An Equality Type Extension, and Applying Telescopes}\label{subsec:telescopeapply}

\begin{figure}[t]
\begin{minipage}[t]{0.57\textwidth}
\begin{alltt}\codefontsize
\hlang[\Turntwo]  (require \modu{dep-lang})

(\defty \defname{=} : [A : Type] [a : A] [b : A] -> Type)
(\defty \defname{refl} : [A : Type] [e : A] -> (= A e e))
(\defred \defname{\match{=}} \pat{(\match{=} pt (refl A t))} ~> \stx{pt})
\end{alltt}
\end{minipage}
\begin{minipage}[t]{0.42\textwidth}
\begin{alltt}\codefontsize
\moduname{dep-lang/eq}
(\deftyrule\,\pat{(\defname{transport} t P p w e)}\expa
  \premss{t}{A} \premcc{w}{A}
  \premcc{P}{(\artt A Type)}
  \premcc{p}{(\er{P} \er{t})}
  \premcc{e}{(= A \er{t} \er{w})}
  --------------
  \concs{(\match{=} \er{p} \er{e})}{(\er{P} \er{w})})
\end{alltt}
\end{minipage}
\figcapspacing
\caption{A \modu{dep-lang} extension for the equality type.}
\label{fig:eq}
\figpostspacing\extrasp
\end{figure}

\fullref[]{fig:eq} presents a module implementing an equality, or identity,
type. The \ttt{transport} rule dictates that for any motive \ttt{P} such that
\ttt{(P a)} holds, eliminating a proof that \ttt{a = b} allows concluding
\ttt{(P b)}. The \ttt{\match{=}} reduction then rewrites to \ttt{\er{pt}} when
the proof is a \ttt{refl} constructor.
The \ttt{=} \ttt{\defty} declaration uses \emph{telescopic}
arguments. While \fullref[]{subsec:telescopecheck} presented our technique of
\emph{checking} telescopes, \emph{applying} telescopic constructors is equally
tricky. This subsection addresses the latter with a novel,
pattern-based substitution technique.

\begin{figure}[t]
\begin{alltt}\codefontsize
(\deftyrule \pat{(\defname{\defty} name : [A : \ktt[1]] \ooo -> \ktt[2])} \expa \moduname{\Turntwo}
  \premcc[\depbic{A}{\ktt[1]}{\Tytt}\ooo ]{\ktt[2]}{\Tytt}
  -----------------------------------
  [\ret \ultt(\deftyrule \pat{(\defname{name} \er{A} ...)} \expa\urtt
       \premcc{\er{A}}{\er[1]{\ki}} \ooo
       -------------------
     \hspace{2pt}\lltt\concs{(\er{name} \er{\er{A}} \ooo)}{\er[2]{\ki}})       \hspace{2pt}\lrtt] \comm{rest of the macro elided}        )
\end{alltt}
\figcapspacing
\caption{(Part of) the implementation of \defty, showing instantiation of telescopes.}
\label{fig:defty}
\figpostspacing\extrasp
\end{figure}

\fullref[]{fig:defty} shows the relevant parts of \ttt{\defty}, which generates
a \ttt{\deftyrule} that uses this technique. \ttt{\defty} first validates the
$\stx{\ktt[1]\ooo\ktt[2]}$ annotations supplied by the programmer, with the
new premise syntax from \fullref[]{subsec:telescopecheck}. The conclusion (a $\ret$
form accommodates emitting top-level forms like definitions) produces the type
rule for \ttt{name} types.  The key is the reuse of the $\pat{\er{A}\;\ooo}$
pattern variables from the premises of the \ttt{\defty} as the pattern
variables of the generated \ttt{\deftyrule}. When the \ttt{name} type
constructor is called, \pat{\er{A}} is bound to the arguments supplied to that
constructor. Since \pat{\er{A}} also binds references in \pat{\er[1]{\ki}\ooo},
however, uses of \stx{\er[1]{\ki}\ooo} in \ttt{name} \emph{automatically} have
\ttt{\er{A}} references replaced with the concrete arguments to the \ttt{name}
type constructor, which is the desired behavior. In other words, we piggyback
on substitutions that the macro system already performs with pattern variables
in templates to instantiate type variables. Further, the technique is safe,
\ie, no variables are captured, thanks to hygiene.

%
%
%
\extrasp\subsection{INTERLUDE: Managing Extensions to a Trusted Core}\label{sec:dep:trustedcore}

Our macro-based approach gives library writers the same power as language
implementers to add new types and rules. Of course, this is dangerous since
they might implement rules incorrectly, changing the trusted core.
\fullref[top]{fig:baddepprog} shows a blatant example. Using \ttt{\asstt}, it
defines \ttt{false=true} as an arbitrary term with type \ttt{(= false true)},
rendering previously unprovable theorems, \eg, that \ttt{(not x)} equals
\ttt{x}, for all \ttt{x}, provable.

\begin{figure}[t]
\begin{alltt}\codefontsize
\hlang[dep-lang]  (require dep-lang/bool dep-lang/eq)\moduname{bad-dep-lang-prog}
(define \defname{false=true} (\asstt \stx{(\er{void})}  \stx{(= false true)})) \comm{should not be able to do this}
(\lm x (\elim{Bool} x (\lm y (= (not y) y)) false=true (sym false=true)))
    \comm{proves \ttt{(\fatt [x : Bool] (= (not x) x))}}
\vspace{4pt}\hlang[dep-lang](require dep-lang/bool dep-lang/eq unsafe/axiom)\moduname{dep-lang-prog-axioms}
(define-axiom \defname{false=true} (= false true))
(print-assumptions (\lm x (\elim{Bool} x (\lm y (= (not y) y)) false=true (sym false=true))))
    \comm{=> Axioms used: \ttt{false=true : (= false true)}}
\vspace{4pt}\hlang[\Turntwo] (provide define-axiom print-assumptions) \moduname{unsafe/axiom}
(\defmac \pat{(\defname{define-axiom} name \ty)}
  \ulltt(\defdef \defname{name} (attach (\asstt \stx{(\er{void})}  \stx{\ty}) \prop{axiom} \stx{name}))\ulrtt)
(\defmac \pat{(\defname{print-assumptions} e)}
  (print "Axioms used: \textasciitilde{}a" (find-axioms (\expatt #`\stx{e})))) \comm{scans \ttt{e} for \textit{axiom} props}
\end{alltt}
%
\figcapspacing
\caption{(top) \modu{dep-lang} program showing danger of extensions. (mid/bot) An axiom library to track extensions.}
\label{fig:baddepprog}
\figpostspacing
\end{figure}

Fortunately, the ability of macros to control syntax can also help \emph{tame}
this power. This subsection shows several possibilities. Any extensions in the
paper may use the following mechanisms.  A first step is to exclude ``unsafe''
extension capabilities, like \ttt{\asstt}, from core
\modu{dep-lang}. \fullref[mid]{fig:baddepprog} shows the same program, but with
``axioms'' marked. Specifically, the program explicitly imports
\ttt{unsafe/axiom} (\fullref[bot]{fig:baddepprog}), which has two constructs:
\ttt{define-axiom} produces an \ttt{\asstt}, but tagged with an extra
\prop{axiom} property;
\ttt{print-assumptions} then scans a term for these marked subterms and
reports them. Using these constructs, programmers can at least know when they
are using unproven axioms. Proof assistants like Coq have a similar feature.

\begin{figure}[t]
\begin{alltt}\codefontsize
\hlang[\Turntwo]   (require rosette) \comm{imports \ttt{z3verify}}\moduname{unsafe/z3}
(\defmac \pat{(\defname{define-axiom/z3} name e \ty)}
  \ulltt(\defmac \defname{name} \stxwhen (z3verify \stx{\ty}) (attach (\asstt \stx{(\er{void})}  \stx{\ty}) \prop{axiom} \stx{name} \prop{z3} \stx{name}))\ulrtt)
\end{alltt}
\begin{alltt}\codefontsize
\vspace{4pt}\hlang[\Turntwo]    (provide (rename [require/report require])) \moduname{dep-lang}
\vspace{2pt}(define (maybe-report-extensions! module-path)
  (when (not (equal? (get-lang module-path) \modu{dep-lang}))
    (print "using extension:" module-path)))
\vspace{2pt}(\defmac \pat{(require/report module-path)}
  \stx{(require module-path) \quad (maybe-report-extensions! module-path)})
\end{alltt}
%
%
\figcapspacing
\caption{(top) solver-aided axioms; (bot) overloading behavior of \ttt{require} and \ttt{provide}.}
\label{fig:verifyrequire}
\figpostspacing\extrasp
\end{figure}

With our framework, we can do more; \fullref[]{fig:verifyrequire} shows two
possibilities. The first (top) is another axiom-definer library; it
provides \ttt{define-axiom/z3}, which is like \ttt{define-axiom}, except it
asks a Z3 solver to verify the theorem before accepting it. Since our
extensions are linguistically supported, not third-party tools, we may use
arbitrary Racket libraries; thus we call on the Rosette language~\cite{rosette}
to help translate to SMT-LIB terms. The resulting term is marked
as \emph{both} \prop{axiom} and \prop{z3axiom}, enabling more fine-grained
classification. (Of course, the solver must now be trusted. We are currently
working to translate Z3 proof scripts back to
\modu{dep-lang} terms.)

A remaining loophole is that users must explicitly use these axiom libraries;
they could just as easily use another library that does not mark
axioms. \fullref[bot]{fig:verifyrequire} shows one way to address this, by
modifying a language's import mechanism.  Specifically, \ttt{require/report}
wraps \ttt{require} so it warns if an imported module is not implemented with
the trusted \modu{dep-lang} core (we can easily error instead of warning if we
want a safe non-extensible language). If \modu{dep-lang} uses
\ttt{require/report} as its only import mechanism, users cannot circumvent it.

\extrasp\subsection{Indexed Inductive Type Families}

Instead of modifying the trusted core with type schemas, most proof assistants
support safe extension with inductively-defined type
families~\cite{Dybjer1994}. In other words, the core is extended just once with
a set of sound, general-purpose rules for defining new types. We
straightforwardly add this capability to \modu{dep-lang}, using the constructs
we have already presented, to
get \modu{dep-ind-lang}. Specifically, \fullref[]{fig:defdataty}
presents \ttt{\defdataty}, which is based on Brady's presentation
of \ttt{TT}~\cite{brady-thesis}. The complete implementation is mere tens of
lines of code,\footnote{Our goal here is to communicate the essence of the
definition clearly; thus we do elide positivity checking and some other
definitions that would clutter the code, but the actual implementation is not
too much longer.} yet it makes
\modu{dep-ind-lang} comparable to the
core of proof assistants like Coq, demonstrating that our macro-based approach
scales to expressive type theories while maintaining convenient notation.

We use a concrete length-indexed list example
to help explain \ttt{\defdataty}:
\begin{alltt}\codefontsize
\hlang[dep-ind-lang]\moduname{list-prog}
(\defdataty \defname{Vec} [A : Type] : (\artt [i : Nat] Type)
  [nil  : (Vec A 0)]
  [cons [k : Nat] [x : A] [xs : (Vec A k)] : (Vec A (S k))])
\end{alltt}
The main source of complexity compared to previous type definitions is that
indexed inductive types distinguish between \emph{parameters} and
\emph{indices} (\ttt{A} and \ttt{i} in the figure).  Parameters
are invariant across the definition while indices may vary.  Thus data
constructor declarations (\ttt{nil} and \ttt{cons}) may reference parameters
\ttt{A} from the type definition, but indices must be specific to each
constructor.

Briefly, the \ttt{\defdataty} macro produces the following
definitions: \ttt{\defty}s to define the type and its data constructors; a
\ttt{\deftyrule} elimination rule; and a \ttt{\defred} reduction rule for the
eliminators. A line-by-line explanation follows.

\begin{figure}[t]
\begin{alltt}\codefontsize
\hlang[\Turntwo]\moduname{dep-ind-lang}
(\deftyrule \pat{(\defname{\defdataty} T [A : \ty[A]] \ooo : \ty  [C x+\ty \ooo : \ty[C]] \ooo)} \expa
 [\depbijerT{A}{\ty[A]}{\er[A]{\ty}}\,\ooo \bijer{T}{(\textPi [A\;:\;\ty[A]]\:\ooo \ty)} \turn \chkjT{\ty}{(\textPi [\er{i}\;:\;\erty[i]]\,\ooo \erty[T])}
                 \hfill\chkjT{(\textPi x+\ty \ooo \ty[C])}{(\textPi [\er{i+x}\;:\;\erty[1]] \ooo (T \erty[2] \ooo))} \ooo ]
 \stxwith \pat{(((\er{i}\sbtt{x} \ooo \er{x}\sbtt{rec}) \ooo) \ooo)} (find-recur \stx{\er{T}} \stx{(([\er{i+x} \erty[1]] \ooo) \ooo)})
 \stxwith \pat{((\erty[2i] \ooo) ...)} (drop-params \stx{((\erty[2] \ooo) \ooo)})
 --------------------------
 [\ret \ultt(\defty \defname{T} : [\er{A} : \er[A]{\ty}] \ooo [\er{i} : \er[i]{\ty}] \ooo -> \erty[T]) \comm{define the type} \hfill\urtt\hspace{8pt}
     \hspace{2pt}(\defty \defname{C} : [\er{A} : \erty[A]] \ooo [\er{i+x} : \erty[1]] \ooo -> (T \erty[2] \ooo)) \ooo \comm{and the data constructors}
     \hspace{2pt}(\deftyrule \pat{(\defname{\elimname} v P m ...)} \comm{define eliminator for T}
       \premss{v}{(T \er{A} \ooo \er{i}\sbtt{v} \ooo)} \comm{target}
       [\turn \stx{P} \expa \pat{\er{P}} \chck \stx{(\Pitt [\er{i}\;:\;\erty[i]]\ooo (\artt (T \er{A} \ooo \er{i} \ooo) Type))}] \comm{motive}
       [\turn \stx{m} \expa \pat{\er{m}} \chck \stx{(\Pitt [\er{i+x}\;:\;\erty[1]]\ooo (\artt (\er{P} \er{i}\sbtt{x}\ooo \er{x}\sbtt{rec})\ooo (\er{P} \erty[2i]\ooo (C \er{A}\ooo \er{i+x}\ooo))))}] \ooo
       ----------------------
       \concs{(\matchname \er{v} \er{P} \er{m} \ooo)}{(\er{P} \er{i}\sbtt{v} \ooo \er{v})})
     \hspace{2pt}(\defred \defname{\match{T}} \comm{define reduction rule for eliminator}
    \lltt  [\pat{(\elim{T} (C \er{A} \ooo \er{i+x} \ooo) \er{P} \er{m} \ooo)} ~> \stx{(\er{m} \er{i+x} \ooo (\matchname \er{x}\sbtt{rec} \er{P} \er{m} \ooo) \ooo)}] \ooo) \hfill\lrtt])
\end{alltt}
\figcapspacing
\caption{The \ttt{\defdataty} form allows \modu{dep-ind-lang} programmers to define inductive datatypes.}
\label{fig:defdataty}
\figpostspacing\extrasp
\end{figure}

\begin{alltt}\codefontsize
\codebullet (\deftyrule \pat{(\defname{\defdataty} T [A : \ty[A]] \ooo : \ty  [C x+\ty \ooo : \ty[C]] \ooo)} \expa
\end{alltt}
%
%
This defines a new language construct named \ttt{\defdataty}. When used,
\ttt{\defdataty} defines a type constructor named \ttt{T} that itself has type
\ttt{(\textPi\,[A : \ty[A]] \ooo\,\ty)}. A colon distinguishes the parameters
from the rest of the type and the \ttt{\ty} part may include indices, as
in the \ttt{Vec} example. The rest of this input pattern specifies the data
constructors \ttt{C \ooo} that produce terms of type \ttt{T}; each \ttt{C} has type
\ttt{(\textPi\;[A : \ty[A]]\;\ooo\;x+\ty\;\ooo\;\ty[C])} where the \ttt{A \ooo}
are the same parameters from \ttt{T}.

%
%
\begin{alltt}\codefontsize
\codebullet [\depbijerT{A}{\ty[A]}{\er[A]{\ty}}\,\ooo\;\bijer{T}{(\textPi [A\;:\;\ty[A]]\,\ooo\;\ty)} \turn \chkjT{\ty}{(\textPi [\er{i}\;:\;\erty[i]]\,\ooo\;\erty[T])}
                 \hfill\chkjT{(\textPi x+\ty \ooo \ty[C])}{(\textPi [\er{i+x}\;:\;\erty[1]] \ooo (T \erty[2] \ooo))} \ooo ]
\end{alltt}
These premises validate the types supplied by a programmer writing a
\ttt{\defdataty}. Since these types may recursively reference the type being
defined, the type environment includes \ttt{T}.  The pattern for the expansion
of \ttt{\ty} includes explicit \patdk{\er{i}} index pattern variables; similarly,
the data constructor type patterns include \patdk{\er{i+x}} variables for the
constructor arguments, which may include indices. Finally, the output of each
data constructor must be of type \ttt{T}, applied to some \ttt{\erty[2] \ooo}.

%
%
\begin{alltt}\codefontsize
\codebullet \stxwith \pat{(((\er{i}\sbtt{x} \ooo \er{x}\sbtt{rec}) \ooo) \ooo)} (find-recur \stx{\er{T}} \stx{(([\er{i+x} \erty[1]] \ooo) \ooo)})
  \stxwith \pat{((\erty[2i] \ooo) ...)} (drop-params \stx{((\erty[2] \ooo) \ooo)})
\end{alltt}
These lines extract subcomponents of the datatype definition that are needed to
define the eliminator and reduction rules. The first \ttt{\stxwith} finds the
recursive arguments of the data constructors; that is, those arguments with
type \ttt{T}, where each \ttt{(\er{i}\sbtt{x} \ooo \er{x}\sbtt{rec})} is a
subset of its corresponding \ttt{(\er{i+x} \ooo)}. The second \ttt{\stxwith}
extracts the index arguments unique to each data constructor, \eg, the index
for \ttt{nil} and \ttt{cons} is
\ttt{0} and \ttt{(S k)}, respectively.

%
%
%
\begin{alltt}\codefontsize
\codebullet (\defty \defname{T} : [\er{A} : \er[A]{\ty}] \ooo [\er{i} : \er[i]{\ty}] \ooo -> \erty[T]) \comm{define the type}
  (\defty \defname{C} : [\er{A} : \erty[A]] \ooo [\er{i+x} : \erty[1]] \ooo -> (T \erty[2] \ooo)) \ooo \comm{and the data constructors}
\end{alltt}
This defines type constructor \ttt{T} and data constructors \ttt{C}. Note that the
latter includes parameters \ttt{A} that were not originally specified with the arguments of \ttt{C}.

%
%
\begin{alltt}\codefontsize
\codebullet (\deftyrule \pat{(\defname{\elimname} v P m ...)} \comm{define eliminator for terms of type T}
    \premss{v}{(T \er{A} \ooo \er{i}\sbtt{v} \ooo)} \comm{target}
    [\turn \stx{P} \expa \pat{\er{P}} \chck \stx{(\Pitt [\er{i}\;:\;\erty[i]] \ooo (\artt (T \er{A} \ooo \er{i} \ooo) Type))}] \comm{motive}
     [\turn \stx{m} \expa \pat{\er{m}} \chck \stx{(\Pitt [\er{i+x}\;:\;\erty[1]] \ooo (\artt (\er{P} \er{i}\sbtt{x} \ooo \er{x}\sbtt{rec}) \ooo (\er{P} \erty[2i] \ooo (C \er{A} \ooo \er{i+x} \ooo))))}] \ooo
    ----------------------
    \concs{(\matchname \er{v} \er{P} \er{m} \ooo)}{(\er{P} \er{i}\sbtt{v} \ooo \er{v})})
\end{alltt}
Defines an eliminator \ttt{\elim{T}} for terms of type \ttt{T}, which has three
arguments: a target \ttt{v}, a motive \ttt{P}, and methods \ttt{m}, one for
each \ttt{C}. This general eliminator definition almost exactly matches its
theoretical presentation in~\cite{brady-thesis}, again showing how \Turntwo
code closely matches its specification. The target \ttt{v} must have type
\ttt{T}. In the pattern for \ttt{v}'s type, the reuse of pattern variables
\patdk{\er{A}} from the premises to \ttt{\defdataty} uses the pattern-based
type instantiation technique introduced in \fullref[]{subsec:telescopeapply}.
Within this elimination rule, any other pattern variables from
\ttt{\defdataty}'s input with references to \ttt{\er{A}}, \eg,
\ttt{\er[i]{\ty}} or \ttt{\er[1]{\ty}}, will automatically be instantiated with
\ttt{v}'s parameters by the macro system. We \emph{do not} use this technique,
however for indices, which are new pattern variables \patdk{\er[v]{i}}.  The
motive \ttt{P} is a function that consumes indices and a value with type
\ttt{T} at those indices, and returns a type for the result of elimination. As
mentioned, \ttt{\er[i]{\ty}} are the types of indexes, but automatically
instantiated with the inferred concrete parameters of target \ttt{v}.

A call to the eliminator must include one method \ttt{m} for each constructor
\ttt{C}. Each method consumes the same inputs as each \ttt{C}, as specified in
the input to \ttt{\defdataty}, as well as one extra argument for each recursive
\ttt{\er{x}\sbtt{rec}}. These latter arguments represent recursive applications
of the eliminators, so their types are specified by the motive \ttt{P}, \ie,
\ttt{(\er{P}\;\er{i}\sbtt{x}\;\ooo \;\er{x}\sbtt{rec})}. The type \ttt{(\er{P}
  \erty[2i] \ooo \;(C \er{A} \;\ooo \;\er{i+x} \ooo))} of each method's result
is also determined by the motive, where the \ttt{\erty[2i]} are the indices
specific to each \ttt{C} constructor.  Finally, the eliminator output calls
reduction rule \ttt{\matchname} to reduce redexes where \ttt{\er{v}} is a
fully-applied constructor. Its type is determined by the motive applied to
\ttt{v} itself.

%
%
\begin{alltt}\codefontsize
 \codebullet (\defred \defname{\match{T}} \comm{define reduction rule for eliminator}
    [\pat{(\elim{T} (C \er{A} \ooo \er{i+x} \ooo) \er{P} \er{m} \ooo)} ~> \stx{(\er{m} \er{i+x} \ooo (\matchname \er{x}\sbtt{rec} \er{P} \er{m} \ooo) \ooo)}] \ooo)])
\end{alltt}
This last definition is a \ttt{\defred} reduction rule
(from~\fullref[]{fig:define-red}) consisting of a series of redexes, one for
each constructor \ttt{C}. It states that elimination of a fully-applied
constructor \ttt{C} reduces to an application of the method for that
constructor, where the recursive arguments to the method are additional
invocations of the eliminator on the recursive constructor arguments. The macro
system's pattern language naturally associates each \ttt{C} with its method
\ttt{m}, again resulting in a concise definition that matches what language
designers write on paper.

%% file: precur.tex
To show that our approach to implementing dependent types scales to realistic
languages, this section presents \modu{Cur}, an extension of
\modu{dep-ind-lang} with features expected in such languages. Specifically, we
add implicit arguments, pattern matching, and recursive top-level function
definitions. To help implement these features, we show how a macro system makes
it straightforward to implement operations like unification and features like
generic methods for types in \Turntwo.

\newcommand{\precurextrasp}{\vspace{0pt}}
\precurextrasp
\subsection{Implicit Arguments and Unification}

\noindent\fullref[]{fig:implicit} shows \ttt{define-implicit}, a form for
declaring implicit arguments (roughly like Coq's ``Arguments'' extension). It
defines \ttt{name\sbtt{abbrv}}, which is equivalent to a given \ttt{name}
function without its first \ttt{n} arguments. The \ttt{define-implicit} form
emits two definitions; the first is a pattern macro that allows omitting
the same arguments in patterns as well. The second, the new \ttt{name\sbtt{abbrv}},
relies on a \Turntwo \ttt{unify} function to compute the omitted arguments. The
\ttt{unify} function consumes constraints in the form of pairs of types that
should be considered equal---here it's the types of its explicit arguments and
the expected type of the whole term, paired with the analogous types from \ttt{name}'s
original function type---and returns a set of substitutions \ttt{\substs} for
the variables in the types that would indeed make the types equal. The second
\ttt{unify} argument is an initial (empty) substitution.

\begin{figure}[t]
\begin{alltt}\codefontsize
\hlang[\Turntwo]    (require \Turntwo/unification) \moduname{\Cur}
(\deftyrule \pat{(\defname{define-implicit} name\sbtt{abbrv} = name #:omit n)}
 \prems{name}{\_}{(\textPi [X : \ty] \ooo \ty[out])}
 \stxwith \pat{(\ty[explicit] \ooo)} (stx-drop \stx{n} \stx{(\ty \ooo)})\quad \stxwith \pat{(X\sbtt{implicit} \ooo)} (stx-take \stx{n} \stx{(X \ooo)})
 ---------------
 [\ret \ultt(define-pattern-m \pat{(\defname{name\sbtt{abbrv}} pat\sbtt{explicit} \ooo)} \stx{(name X\sbtt{implicit} \ooo pat\sbtt{explicit} \ooo)})\hfill\urtt\hspace{8pt}
     \hspace{2pt}(\deftyrule \pat{(\defname{name\sbtt{abbrv}} arg\sbtt{explicit} \ooo)} \chck \pat{\ty[expect]} \expa
       \prems{arg\sbtt{explicit}}{\er[explicit]{arg}}{\ty[arg]} \ooo
       \stxwith \pat{\substs} (unify \stx{([\ty[expect] \ty[out]] [\ty[arg] \ty[explicit]] \ooo)} \stx{()})
       \stxwith \pat{(\er[implicit]{arg} \ooo)} (lookup \stx{X}\sbtt{implicit} \stx{\substs})
       ------------------------
    \hspace{1.5pt}\lltt  [\ret \stx{(name \er[implicit]{arg} \ooo \er[explicit]{arg} \ooo)}])\hfill\lrtt])
\end{alltt}
\figcapspacing
\caption{\modu{Cur} extension for declaring implicit arguments.}
\label{fig:implicit}
\figpostspacing
\precurextrasp
\end{figure}

%
%
%
%
\newcommand{\dotpat}{\hspace{-2pt}.\hspace{-2pt}}
\begin{figure}[t]
\begin{alltt}\codefontsize
(define (\defname{unify} constraints\sbtt{\ty=\ty} \substs)\moduname{\Turntwo/unification}
 (syntax-parse constraints\sbtt{\ty=\ty}
  [\pat{()} \substs] \comm{(1) base case} \quad [\pat{([\ty x] \!.\! rst)} (unify \stx{([x \ty] . rst)} \substs)] \comm{(2) swap id to lhs}
  [\pat{([x \ty] \dotpat rst)} \stxwhen (in? \stx{x} (dom \substs)) (unify \stx{([}(lookup \stx{x} \substs) \stx{\ty] \dotpat rst)} \substs)] \comm{(3) conflict}
  [\pat{([x \ty] \dotpat rst)} \stxwhen (not (occurs \stx{x} \stx{\ty})) \comm{(4) elim}
   (unify (subst \stx{\ty} \stx{x} \stx{rst}) \stx{([x \ty] }(subst\sbtt{rng} \stx{\ty} \stx{x} \substs)\stx{)})]
  [\pat{([\ty[1] \ty[2]] \dotpat rst)} \stxwhen (\ty= \ty[1] \ty[2]) (unify \stx{rst} \substs)] \comm{(5) delete}
  [\pat{([(\er{\app} C\sbtt{1} \ty[1]\ooo) (\er{\app} C\sbtt{2} \ty[2]\ooo)] \dotpat rst)} \stxwhen (and (= \stx{C\sbtt{1}} \stx{C\sbtt{2}}) (len= \stx{(\ty[1] \ooo)} \stx{(\ty[2] \ooo)}))
   (unify \stx{([\ty[1] \ty[2]] \ooo . rst)} \substs)] \comm{(6) decompose}
  [\pat{([(\er{\lm} x\sbtt{1} e\sbtt{1}) (\er{\lm} x\sbtt{2} e\sbtt{2})] \dotpat rst)} (unify \stx{([e\sbtt{1} }(subst \stx{x\sbtt{1}} \stx{x\sbtt{2}} \stx{e\sbtt{2}})\stx{] \dotpat rst)} \substs)] \comm{(7) HO case}
  [\pat{([\ty[1] \ty[2]] \dotpat rst)} (tyerror "could not unify \stx{\ty[1]} and \stx{\ty[2]}")])) \comm{(8) error}
\end{alltt}
\figcapspacing
\caption{\Turntwo basic unification function.}
\label{fig:unify}
\figpostspacing
\precurextrasp
\end{figure}

\fullref[]{fig:unify} shows a basic unification function as a \Turntwo library;
it is roughly the well-known \citet{martelli-montanari82} algorithm, with an
extra higher-order case for binding forms. We show it to demonstrate how a
macro system's syntax manipulation and handling of binding makes it
straightforward to implement operations like unification, since the
implementation cases more or less correspond to the algorithm's
specification. The eight cases may be summarized as (in order of
implementation): (1) base case, which returns the accumulated substitution
\ttt{\substs}; (2) swaps identifiers to the left side; (3) handles conflicting
substitutions for a variable by adding a constraint; (4) eliminates a
constraint by adding a substitution for \ttt{x} to \ttt{\substs}, replacing
\ttt{x} with \ttt{\ty} in the existing substitutions and constraints; (5) drops
a constraint if the types are equal; (6) decomposes constructor applications
into constraints for its arguments; (7) a higher-order case that adds a new
constraint for the bodies; and (8) an error case when a constraint has types
that are not equal.

%
%
\precurextrasp
\subsection{Dependent Pattern Matching and Generic Type Methods}\label{sec:cur:match}

Explicit eliminators are unwieldly to use; most programmers prefer pattern
matching instead. \fullref[]{fig:match} sketches a dependent pattern matcher
that is sugar for the underlying eliminator. (This basic matcher's goal is to
illustrate \Turntwo's generic interface. We are exploring more sophisticated
translations, \eg,~\citet{goguen2006}.) The key to \ttt{match} is
determining which eliminator to use. More specifically, for any given type,
we need the original datatype definition. In the implementation of \ttt{match},
a generic \ttt{get-datatype-def} function returns this information. This
function must behave differently depending on its argument type, however, and
thus its implementation relies on a generic method interface for types in
\Turntwo. Once \ttt{match} has the original datatype definition, it may
generate the equivalent eliminator term by: (1) computing the eliminator name,
(2) checking that the case patterns are complete, and (3) converting the
case patterns to eliminator methods by adding the recursive arguments.

\begin{figure}[t]
\begin{alltt}\codefontsize
(\deftyrule \pat{(\defname{match} e #:as x #:with-indx i\ooo #:in \ty[in] #:return \ty[out] case\ooo)}\,\expa\moduname{Cur}
  \premcc{e}{\ty[in]}
  \stxwith \pat{def} (get-datatype-def \stx{\ty[in]}) \quad \stxwith \pat{elim-name} (get-elim-name \stx{def})
  \stxwhen (cases-complete? \stx{(case \ooo)} \stx{def})
  [\ret \stx{(elim-name \er{e} (\lm i \ooo x \ty[out]) (case->method case def) \ooo)}])
\vspace{4pt}(\defmac \pat{(\defname{case->method} [(C x ...) body] def)}
  \stxwith \pat{(x\sbtt{rec} ...)} (get-x\sbtt{rec} \stx{(C x ...)} \stx{def})
  \stx{(\lm x ... x\sbtt{rec} ... body)})
\end{alltt}
\figcapspacing
\caption{A pattern matcher for \modu{Cur}.}
\label{fig:match}
\figpostspacing
\precurextrasp
\end{figure}

\begin{figure}[t]
\begin{alltt}\codefontsize
\comm{(part of) \Turntwo generic API}\moduname{\Turntwo/typedef}
(\deftyrule \pat{(\defname{\defty} tyname ... #:implements methname meth \ooo)}  \comm{\ooo}
 \lltt(define \defname{\er{tyname}} (hash-table \stx{methname} \stx{meth \ooo}))\lrtt)
(\defmac \pat{(\defname{define-generic-type-method} methname)}
 \ultt(\deftyrule \pat{(\defname{methname} \ty)} \expa             \urtt
   \premc{\ty}{(\er{\app} \er{tyname} \_\ooo)}{\Tytt}
 \lltt [\ret (dict-ref (get-dict \stx{\er{tyname}}) \stx{methname})])\lrtt)
\vspace{4pt}(define-generic-type-method \defname{get-datatype-def}) \comm{Use of \Turntwo generic API}\moduname{Cur}
(\deftyrule \pat{(\defname{define-datatype} \ooo)}  \comm{\ooo}
 \lltt(define-type \ooo #:implements get-datatype-def (\lm (ty) \stx{<the entire datatype def>}))\lrtt)
\end{alltt}
\figcapspacing
\caption{(top) \Turntwo's generic interface for types, (bot) example use of the interface}
\label{fig:generic}
\figpostspacing
\precurextrasp
\end{figure}

\fullref[top]{fig:generic} sketches how \Turntwo's generic type method
interface works. First, \ttt{\defty} from section~\ref{subsec:patexpanders} is
modified to additionally emit one more definition: a method table that is
available during macro expansion. Then a programmer, using
\ttt{define-generic-type-method}, may define a generic rule that, based on a
given type, looks up the method in that type's table and dispatches to it. It
can do this because the internal name of the type \stx{\er{tyname}} is also the
name of the method table. There's no name conflict because the names are bound
at different phases of macro expansion.

\fullref[bot]{fig:generic} shows a usage of this generic interface to implement
\ttt{get-datatype-def}, the generic method used in~\fullref[]{fig:match} that
returns the original datatype definition. A programmer first declares the
generic method with \ttt{define-generic-type-method}. Then, the \ttt{\defty}
generated by \ttt{\defdataty} uses the extra \ttt{\#:implements} option to
define type-specific versions of the function. In this case, it just returns
the entire input to \ttt{\defdataty}.

\precurextrasp
\subsection{Recursive Function Definitions} \label{sec:precur:recur}

Realistic programming languages allow programmers to write recursive top-level
pattern-matching function definitions. Some languages might define such a construct
as sugar over {\lm}s which, when combined with \ttt{match} from
section~\ref{sec:cur:match}, gets part of the way there, \eg:
\begin{alltt}\codefontsize
(\deftyrule \pat{(\defname{define\sbtt{bad}} f [x : \ty] \ooo : \ty[out] body)} \expa
  \premc[\bi{x}{\ty}\ooo \bi{f}{(\textPi [x : \ty] \ooo \ty[out])}]{body}{\er{body}}{\ty[out]}
  ------------
  [\ret \stx{(\er{define} \er{f} (\er{\lm} (\er{x} \ooo) \er{body}))}])
\end{alltt}
Function \ttt{f} may call itself in its body because (1) \ttt{f} is
added to the type environment, and (2) \ttt{\er{define}} in the target language
allows recursive definitions. This falls short for \modu{Cur}, however,
for several reasons. First, it allows defining non-terminating functions like
\ttt{(define loop [n : Nat] : Nat (loop n))}. Second, our reduction rules fire
too eagerly so that even supposedly terminating functions will not
terminate. For example, imagine the following \ttt{Nat} function:
\begin{alltt}\codefontsize
  (define\sbtt{bad} \defname{f} [n : Nat] : Nat (match n [Z Z] [(S m) (f m)]))
  \comm{\texttt{(f m)} => \texttt{(match m [Z Z] [(S m\sbtt{2}) (f m\sbtt{2})])}}
  \comm{\texttt{     } => \texttt{(match m [Z Z] [(S m\sbtt{2}) (match m\sbtt{2} [Z Z] [(S m\sbtt{3}) (f m\sbtt{3})])])} => ...}
\end{alltt}
This function seems to terminate because \ttt{f} is only called recursively
with a smaller argument. If \ttt{f} is sugar for a {\lm}, however, the
application \ttt{(f m)} in the body can always reduce with \ttt{\betar} and
this will produce infinite applications of \ttt{f}. Instead, recursive
functions should not reduce until they are applied to concrete
constructors, \ie, each pattern case should be a new reduction definition.

\fullref[]{fig:defrecmatch} presents \ttt{\defrecmatch}, which defines
top-level pattern-matching recursive functions. It is simplified to one
argument \ttt{x}, and non-dependent, non-nested patterns, and we do not show
features like inaccessible patterns, in order to focus on the termination
checking (we plan to eventually support Agda-style matching,
\eg,~\cite{coquand1992,cockx14}). The first \ttt{\stxwith} computes binders and
their types from each pattern with a generic method \ttt{pat->ctxt}. The second
creates the expected type of each case body by replacing \ttt{x} in
\ttt{\ty[2]} with the case pattern.

The third and fourth \ttt{\stxwith} tag
marks some of the types with extra syntax properties, enabling termination
checking. Specifically, the function's argument type \ttt{\ty} is marked as a
``rec arg'' while the types returned by \ttt{pat->ctxt} are marked ``rec ok''
because they are ``smaller'' arguments. The last premise type checks the bodies
of each case in a type context with: (1) the original input \ttt{x}, (2)
binders \ttt{x\sbtt{pat}} from the patterns, and (3) the function \ttt{f}
itself. Note this \ttt{f}'s input type is now \ttt{\ty[rec]}; similarly the
type of \ttt{x\sbtt{pat}} is the ``marked'' \ttt{\ty[recok]}. But the original
argument \ttt{x} has unmarked type \ttt{\ty}.  A \ttt{\#:where} directive
following the premise explains how these extra syntax properties are
used. Specifically, the (overloadable) type equality function \ttt{\ty=}
\emph{is changed for just the duration of checking the bodies}, in the
following way. Two types are equal if they were equal with the old
\ttt{\ty=}. Additionally, if the second type is a ``rec arg'', then the first
type must be ``rec ok'', otherwise we raise a type error. In this way, only
terminating recursive function definitions are allowed. Note that in a body,
attempting to apply \ttt{f} to its original input \ttt{x} would fail because
\ttt{x} is not marked ``rec ok''.
A \ttt{\defrecmatch} use emits two
definitions, a \ttt{\deftyrule} defining \ttt{f} and a \ttt{\defred} specifying
how to reduce applications of \ttt{f}. Specifically, an applied \ttt{f} is only
reduced if its argument is a concrete value that matches one of the input
patterns from its definition. In this way, recursive functions avoid the
non-terminating reductions described earlier.

\begin{figure}[t]
\begin{alltt}\codefontsize
(def (mk-recarg \ty) \att{\ty}{rec-arg}{true}) (def (mk-ok \ty) \att{\ty}{rec-ok}{true})\moduname{Cur}
\vspace{4pt}(\deftyrule \pat{(\defname{\defrecmatch} f [x : \ty] : \ty[2] [pat bod] \ooo)}
 \stxwith \pat{(([x\sbtt{pat} \ty[pat]] \ooo) \ooo)} \stx{(}(pat->ctxt \stx{pat} \stx{\ty}) \stx{\ooo)} \hfill\stxwith \pat{(\ty[out]\ooo)} \stx{(}(subst \stx{pat} \stx{x} \stx{\ty[2]) ...)}
 \stxwith \pat{\ty[rec]} (mk-recarg \stx{\ty})\hfill\stxwith \pat{((\ty[recok] \ooo) \ooo)} \stx{((}(mk-ok \stx{\ty[pat]}) \stx{\ooo) \ooo)}
 [ \bijer{x}{\ty} \bijer{x\sbtt{pat}}{\ty[recok]} \ooo \bijer{f}{(\textPi [x : \ty[rec]] \ty[2])} \turn \chkjer{bod}{\ty[out]} \ooo
   #:where \boxed{\ty=}\,(\er{\lm}\,(\ty[1] \ty[2])\,(and (\boxed{\ty=\sbtt{OLD}} \ty[1] \ty[2])
                \hfill(or (not (rec-arg? \ty[2])) (rec-ok? \ty[1]) (err "nonterminate"))))]
 -------------
 [\ret \ultt(\deftyrule (\defname{f} e) \expa                 \urtt
       \premcc{e}{\ty}
       ---------------
       \concs{(f-eval \er{e})}{\ty[2]})
     \lltt(\defred \defname{f-eval} [\pat{(\er{f} pat)} ~> \stx{\er{bod}}] \ooo)\lrtt])
\end{alltt}
\figcapspacing
\caption{A new recursive definition form for \modu{Cur}, with termination checking.}
\label{fig:defrecmatch}
\figpostspacing
\precurextrasp
\end{figure}

%
%
\precurextrasp
\subsection{Sized Types: Implementing Auxiliary Type Systems}

\fullref[]{sec:precur:recur}'s termination check roughly corresponds
to~\citet{gimenez1995}'s syntactic guards in many proof assistants. This
conservative analysis, however, rejects some valid programs, \eg, division via
subtraction, because ``non-increasingness'' of arguments does not propgate
through function calls:
\begin{alltt}\codefontsize
(\defrecmatch \defname{minus} [n : Nat] [m : Nat] : Nat
  [Z \_ => n] [\_ Z => n]
  [(S n-1) (S m-1) => (minus n-1 m-1)])
(\defrecmatch \defname{div\sbtt{rejected}} [n : Nat] [m : Nat] : Nat
  [Z \_ => n]
  [(S n-1) m => (S (div\sbtt{rejected} (minus n-1 m) m))]) \comm{syntactic termination check rejects recursive call}
\end{alltt}

\begin{figure}[t]
\begin{alltt}\codefontsize
(define \defname{inc-sz} [\pat{(< i)} \stx{i}] [else (fresh)]) \quad(define (\defname{dec-sz} sz) \stx{(< }sz\stx{)})\moduname{Cur/sizedtypes}
(define (\defname{get-sz} \ty) (or \dett{\ty}{sz} INF)) (define (\defname{add-sz} \ty sz) \att{\ty}{sz}{sz})
\vspace{4pt}(\deftyrule \pat{(\defname{lift-datatype} TY)}
 \stxwith \pat{df} (get-datatype-def \stx{TY}) \stxwith \pat{(C \ooo)} (get-datacons \stx{df}) \stxwith \pat{(\ty[C] \ooo)} (get-\ty[C] \stx{df})
 [\ret \ultt(\deftyrule \pat{(\defname{C\sbtt{sz}} arg)} \expa              \urtt
       \stxwith \pat{sz} (inc-sz (get-sz \stx{arg}))
        --------------
        [\turn \stx{(C arg)} \synth (add-sz \stx{\ty[C]} \stx{sz})]) \ooo
      (define-instance \defname{C\sbtt{sz}} (pat->ctxt pat ty)
        \stxwith \pat{([x \ty] \ooo)} (pat->ctx (subst \stx{C} \stx{C\sbtt{sz}} pat) ty)
    \lltt   \stx{([x }(dec-sz \stx{\ty})\stx{] \ooo)}) \ooo              \lrtt])
\vspace{4pt}(\deftyrule \pat{(\defname{def/rec/match\sbtt{sz}} f [x : \ty #:sz i] : \ty[out] #:sz j [pat bod] \ooo)}
 \stxwith \pat{\ty[i]} (add-sz \stx{\ty} \stx{i}) \quad \stxwith \pat{\ty[out/j]} (add-sz \stx{\ty[out]} \stx{j})
 \stxwith \pat{(([x\sbtt{pat} \ty[pat]] \ooo) \ooo)} \stx{(}(pat->ctxt \stx{pat} \stx{\ty[i]}) \stx{\ooo)} \comm{\texttt{\ty[pat] \ooo} has size \ttt{(< i)}}
 \stxwith \pat{\ty[<i]} (add-sz \stx{\ty} \stx(< i)) \quad \stxwith \pat{\ty[out/<j]} (add-sz \stx{\ty[out]} \stx{(< j)})
 [ \bijer{x}{\ty[i]} \bijer{x\sbtt{pat}}{\ty[pat]} \ooo \bijer{f}{(\textPi [x : \ty[<i]] \ty[out/<j])} \turn [\stx{bod} \expa \pat{\er{bod}} \chck \stx{\ty[out/j]}]  \ooo
   #:where \boxed{\ty=} (\er{\lm} (\ty[1] \ty[2]) (and (\boxed{\ty=\sbtt{OLD}} \ty[1] \ty[2]) (sz-ok? (get-sz \ty[1]) (get-sz \ty[2])))) ]
 -------------
 [\ret \ultt(\deftyrule (\defname{f} e) \expa                 \urtt
       \premcc{e}{\ty} \stxwith \pat{i} (get-sz \stx{e})
       ---------------
       [\turn \stx{(f-eval \er{e})} \synth (add-sz \stx{\ty[out]} \stx{j})])
     \lltt(\defred \defname{f-eval} [\pat{(\er{f} pat)} ~> \stx{\er{bod}}] \ooo)\lrtt])
\vspace{4pt}(define (\defname{sz-ok?} sz1 sz2) \comm{true when sz1 \texttt{<=} sz2}
 (or (INF? sz2) (syntax-parse (sz1 sz2)
                 [\pat{(x y)} (and (id? x) (id? y) (id=? \stx{x} \stx{y}))]\hfill[\pat{((< x) (< y))} (sz-ok? \stx{x} \stx{y})]
                 [\pat{((< x) y)} (sz-ok? \stx{x} \stx{y})]   \hfill[else (err "non-terminating!")] )))
\end{alltt}
\figcapspacing
\caption{A library for \modu{Cur} that adds sized types.}
\label{fig:sized}
\figpostspacing
\end{figure}

\noindent{}An alternative is sized types~\cite{hughes96sizedtypes}, a more
expressive termination analysis, but adding them is tricky due to its
invasive nature on both a language's implementation and its
usability. Concretely, they typically require threading an extra size
argument into every type and term (see \cite{abel12sized}'s work, which
inspired our ideas, for details) cluttering code and complicating operations
like type equality. This section shows an experimental sized types library that
potentially reaps the benefits while minimizing the negatives. More
specifically, with macros and syntax properties, we can add ``auxiliary'' type
systems, like sized types, that \emph{operate in parallel} to the main one. The
extra types are used when needed, \eg, checking termination, and ignored when
not, \eg, type equality.

\fullref[]{fig:sized} shows the essence of our library. ``Sized'' types are
plain types annotated with a ``sz'' property, where the size property can be
either an arbitrary identifier \ttt{i}, or \ttt{(< sz)} where \ttt{sz} is
another size property. With the extensibility afforded by macros, we only need
to overload a few features. Specifically, the library consists of two main
forms: \ttt{lift-datatype}, which lifts an existing datatype definition to be
sized, and \ttt{def/rec/match\sbtt{sz}}, which reimplements
\ttt{\defrecmatch} from~\fullref[]{fig:defrecmatch} except with sized types for
termination analysis. Here is the previous \ttt{div} example:
\begin{alltt}\codefontsize
\hlang[Cur]    (require cur/sizedtypes)    (lift-datatype Nat)
(def/rec/match\sbtt{sz} \defname{minus\sbtt{sz}} [n : Nat #:sz i] [m : Nat] : Nat #:sz i \comm{\texttt{minus\sbtt{sz}} is non-increasing in size}
  [Z\sbtt{sz} \_ => n]\quad [\_ Z\sbtt{sz} => n]
  [(S\sbtt{sz} n-1) (S\sbtt{sz} m-1) => (minus\sbtt{sz} n-1 m-1)])
(def/rec/match\sbtt{sz} \defname{div\sbtt{sz}} [n : Nat #:sz i] [m : Nat] : Nat #:sz i
  [Z\sbtt{sz} \_ => n]
  [(S\sbtt{sz} n-1) m => (S\sbtt{sz} (div\sbtt{sz} (minus\sbtt{sz} n-1 m) m))]) \comm{sized type termination accepts}
\end{alltt}
\ttt{lift-datatype} takes a type \ttt{TY},
previously defined with \ttt{\defdataty}, and defines sized wrappers
\ttt{C\sbtt{sz}} for each unsized constructor \ttt{C}. To simplify understanding,
we show each \ttt{C} with only one argument; an actual
implementation would have to choose the ``decreasing'' argument. Each
\ttt{C\sbtt{sz}} constructor adds a size to the type of an applied \ttt{C} term
that is the size of its argument ``incremented'', where incrementing a size
either removes a \ttt{<} or generates a fresh id. Dually, \ttt{lift-datatype}
overloads the generic \ttt{pat->ctxt} for \ttt{C\sbtt{sz}} so that the pattern
binders have type that is ``decremented''. This new \ttt{pat->ctxt} is used by
the new \ttt{def/rec/match\sbtt{sz}}.

Except for the different termination analysis, the new
\ttt{def/rec/match\sbtt{sz}} is roughly the same as its predecessor. To
implement termination via sized types, the new definition form requires size
annotations on its types, which are then used while type checking the body of
each case. Observe that when type checking the bodies, the type for \ttt{f}
in the type environment requires an argument that is sized less than the
original argument size \ttt{i}. This smaller argument will typically come from
the result of the new \ttt{pat->ctxt} (generated by \ttt{lift-datatype}). Like
previous \ttt{def/rec/match}, \ttt{\boxed{\ty=}} is overloaded, this time to
ensure that sized types satisfy a \ttt{sz-ok?} predicate, which enforces that
its first argument has size less than or equal to its second. The \ttt{sz-ok?}
function special-cases an \ttt{INF} size, which is assigned to unlifted types,
for when we do not care about sizes.  Most importantly, the size annotation on
the output, which may reference sizes of the input arguments, allows declaring
that functions like \ttt{minus} have non-increasing size. This allows size
information to propagate across function calls to enable \ttt{div}, and even
higher-order cases like ``rose trees'' (see~\cite{abel10sized}).

%% file: cur.tex
\newcommand{\racket}{\ttt}
\newcommand{\todo}[1]{TODO: #1}
\newcommand{\curextrasp}{\vspace{-5pt}}

Even with~\fullref[]{sec:precur}'s extensions, it is still tedious to program
and prove with \modu{Cur}. To make proving practical, proof assistants
typically layer companion DSLs on top of their core. By building with macros
from the beginning, we already have a framework in which both language implementers
and users can easily build such DSLs. They may even build metaDSLs to build
their DSLs, as advocated by language-oriented programming. Best of all, any new DSLs are
linguistically integrated with \Cur, instead of operating as third-party
preprocessors. This section presents three DSLs: \modu{Olly}, for modeling
programming languages; \modu{\ntac}, a tactic language for scripting proofs;
and \modu{\mntac}, a metaDSL used to implement \modu{\ntac}. All these DSLs
elaborate to core \modu{Cur} before type checking; thus we can extend the
functionality of our language yet keep the trusted base small.



\begin{figure}[t]
\begin{minipage}[t]{0.484\textwidth}
\begin{alltt}\codefontsize
\hlang[cur] (require cur/olly) \moduname{olly-prog}
(define-language \defname{stlc} \#:vars (x)
 \#:coq-out "stlc.v" \#:latex-out "stlc.tex"
 val (v) ::= true false unit
 type (A B) ::= boolty unitty (-> A B) (* A A)
 trm (e) ::= x v (lm (\#:bind x : A) e) (ap e e)
 \hspace{3pt}(cons e e) (let (\#:bind x \#:bind x) = e in e))
\end{alltt}
\end{minipage}
\begin{minipage}[t]{0.51\textwidth}
\begin{alltt}\codefontsize
(\defdataty \defname{stlc-trm} : Type
\hspace{3pt}(Var->stlc-trm Var\;:\;stlc-trm)
\hspace{3pt}(stlc-val->stlc-trm stlc-value\;:\;stlc-trm)
 (stlc-lm Var stlc-type stlc-trm\;:\;stlc-trm)
\hspace{3pt}(stlc-ap stlc-trm stlc-trm\;:\;stlc-trm)
\hspace{3pt}(stlc-cons stlc-trm stlc-trm\;:\;stlc-trm)
 (stlc-let\,Var\,Var\,stlc-trm\,stlc-trm\,:\,stlc-trm))
\end{alltt}
\end{minipage}
\figcapspacing
\caption{(l) STLC with \modu{Olly}, a \modu{Cur} notation extension; (r) \modu{Olly}-generated \modu{Cur} datatype}
\label{fig:olly-stlc}
\figpostspacing\curextrasp
\end{figure}

%
%
%
\noindent\textbf{\modu{Olly}} is an Ott-inspired~\cite{sewell:2007} DSL for
modeling programming languages in \modu{Cur}. Specifically, programmers write
BNF or inference rule notation to specify language syntax and relations,
respectively, and \modu{Olly} generates the \modu{Cur} inductive type
definitions; \LaTeX{} or Coq extraction is also supported.
\fullref[l]{fig:olly-stlc} shows the STLC in BNF using \modu{Olly}. Optional
\ttt{\#:bind} annotations specify binding positions in the grammar;
here \ttt{cons} creates pairs and \ttt{let} eliminates them, thus the latter binds two
names. \olly generates an inductive datatype for each non-terminal in the
grammar; \fullref[r]{fig:olly-stlc} shows \ttt{trm}, whose constructor names
are derived from the specification. Extra constructors, \eg,
\ttt{Var->stlc-trm}, allow converting from the other
non-terminals. Internally, \ttt{define-language} uses an intermediate data
structure, which is converted to \modu{Cur}, Coq, \LaTeX{}, and other outputs.
Unlike Ott and other external tools, \modu{Olly} is a user-written library and
is supported linguistically; thus programmers may use \modu{Olly} forms
alongside normal \modu{Cur} code rather than switch to external tools,
demonstrating how our macros-based approach supports tailoring all aspects of a
proof assistant to specific domain, from the object theory to the syntax.

%


%
%
%
\noindent\textbf{Tactic systems} are a popular tool to enable interactive, command-based
construction of proof terms in proof assistants and our macro-based approach
naturally provides all the capabilities required to build one:
pre-type-checking general purpose computation, traversal and pattern matching
of language terms, interesting elaboration system data structures for
manipulating proof states, an API to the object language to type check and
evaluate terms while constructing proofs, interactivity, and syntactic
integration into the language. Even better, we may abstract over these
low-level features with a meta DSL, to implement the tactics concisely and
intuitively.

We present \modu{\ntac}, a tactic language for \modu{\Cur}. It
tracks intermediate hole-embedded proof terms, and subgoals and contexts
corresponding to those holes, as a tree. Further, a zipper navigates
and focuses on subgoals in this proof tree. Concretely, an \modu{\ntac} tactic
is a macro that when invoked, produces a function that transforms zipper data
structure instances. Here is a trivial \modu{ntac} proof:
\begin{alltt}\codefontsize
\hlang[cur] \quad(require cur/ntac) \moduname{example-theorem}
(define \defname{id} (ntac (\fatt (A : Type) (a : A) A) (intros A a) assumption))
\end{alltt}
The first argument to \ttt{ntac} is a goal theorem; the rest are tactic
invocations, \ie, a script that builds the proof term. When invoked,
\ttt{ntac} builds a tree node with the given goal and a hole term, and
creates a zipper with that initial node as the focus. Each subsequent tactic
invocation transforms the zipper and proof tree with nodes that gradually fill
the hole(s). After the script completes, the holeless tree is converted to a
complete proof term, which is the result of the \ttt{ntac} call. Thus, an \ttt{ntac}
invocation may be used in any expression position, \eg, bound to the name
\ttt{id} using \ttt{define}.
A \Cur tactic manipulates the proof zipper but a tactic programmer should not
have to explicitly manage this proof state. \fullref[l]{fig:definetactic} shows
\ttt{define-tactic}, from a \modu{\mntac} language for writing tactics, which
abstracts over low-level details. Specifically, a \ttt{define-tactic}
definition consists of a series of cases, each starting with a pattern
dictating usage syntax of the tactic, and an optional pattern matching the
current goal. The body of each case must return a term, possibly with holes,
whose type matches the current goal. Programmers use \ttt{\node} to
construct this partial term, where optional \ttt{\#:where} declarations, of
shape \ttt{\stx{[x : \ty] \ooo}\,\turn\;?HOLE\,:\,\stx{<subgoal>}}, specify new
subgoals to prove, each corresponding to a \ttt{?HOLE} name referenced in the
first argument of \ttt{\node}. They may also use implicitly bound variables if
they wish to directly access parts of the proof state, \eg, \ttt{\$goal},
\ttt{\$ctx}, \ttt{\$pt}, \ttt{\$ptz}, for the goal, context, proof tree, and
zipper instance, respectively.

\begin{figure}[t]
\begin{minipage}[t]{0.49\textwidth}
\begin{alltt}\codefontsize
\hlang[\mntac] \comm{\ttt{define-tactic} usage pattern}
(define-tactic \defname{tactic-name}
 [\pat{<usage pat>} \#:current-goal \pat{<goal pat>}
  \comm{\ooo}
  \comm{implicit bindings: \ttt{$ctx}, \ttt{$ptz}, \ttt{$pt}, \ttt{$goal}}
  \comm{\ooo}
  (\node \stx{<new partial term with ?HOLEs>}
   \#:where
    [\stx{x : \ty \ooo} \turn ?HOLE1 : \stx{<subgoal1>}] \ooo)]
   \ooo)
\end{alltt}
\end{minipage}
\begin{minipage}[t]{0.5\textwidth}
\begin{alltt}\codefontsize
\hlang[\mntac]\moduname{\ntac}
(define-tactic \defname{intro}
 [\pat{(\_ y)} #:current-goal \pat{(\textPi (x : P) \ty)}
  (\node \stx{(\lm (y : P) ?H)}
    \#:where \stx{y : P} \turn ?H : (subst \stx{y} \stx{x} \stx{\ty}))]
 [\pat{\_} #:current-goal \pat{(\textPi (x : P) \ty)}
  (\node \stx{(\lm (x : P) ?H)} \#:where \stx{x : P} \turn ?H : \stx{\ty})])
\vspace{2pt}(define-tactic \defname{assumption}
 [\pat{\_} (\node (ctx-find \$ctx (\lm t (\ty= t \$goal))
           \#:fail "no assumption \$goal"))])
\end{alltt}
\end{minipage}
\vspace{6pt}
\begin{alltt}\codefontsize
(define-tactic \pat{(\defname{inversion} H #:with-names H\sbtt{0} ...)}    \stxwith \pat{\ty[H]} (typeof \stx{H})\moduname{\ntac}
 \stxwith \pat{((A \ooo) (i \ooo) [C x \ooo x\sbtt{rec} \ooo : \ty\sbtt{C}] ...)} (get-datatype-def \stx{\ty[H]})
 (\node \stx{(elim H (\lm i \ooo H \$goal) (\lm x \ooo x\sbtt{rec} \ooo \$pf) \ooo)} #:with-subgoals
  \stx{(}(stx-parse (unify+prove (get-indxs \stx{\ty[H]}) (get-indxs \stx{\ty[C]}))
    [\pat{([=pf\,:\,=thm]\,\ooo)}\,(\node \stx{((\lm [H0\,:\,=thm]\,\ooo\,?HO)\,=pf\,\ooo)}\,#:where\,\stx{[H0\,:\,=thm]\,\ooo}\,\turn\,?HO\,:\,\$goal)]
    [\pat{pf-of-False} \comm{unify fail} (\node \stx{(elim\sbtt{False} pf-of-False (\lm \_ \$goal))})])\stx{ \ooo)}
\end{alltt}
\figcapspacing
\caption{(l) \ttt{define-tactic} usage; (r) implementation of \ttt{intro} and \ttt{assumption}; (bot) \ttt{inversion} tactic}
\label{fig:definetactic}
\figpostspacing\curextrasp
\end{figure}

\fullref[r]{fig:definetactic} shows the implementation of the two tactics used
in the \ttt{id} theorem above. The \ttt{intro} tactic (only
single-variable cases are shown) has two cases; the first is invoked when
given an identifier \ttt{y}, when the goal has shape \ttt{(\textPi \,[x : P]
  \ty)}. It fills the current hole with a \ttt{\lm} binding \ttt{[y : P]},
which has a new hole \ttt{?H} in its body. It then specifies that a new subgoal
for \ttt{?H}, in context where \ttt{y} has type \ttt{P}, is \ttt{\ty} except
with \ttt{x} replaced by the argument \ttt{y}. The second \ttt{intro} case
has no argument; instead, it directly uses \ttt{x} from
the goal. The \ttt{assumption} tactic also has no argument; it
searches the current context, bound to \ttt{\$ctx}, for a variable with type
matching the current goal. If it's successful, it fills the current hole with
the variable; otherwise it raises an error. No \ttt{\#:where} argument for
\ttt{\node} is necessary here because the resulting proof term has no holes.

\modu{\mntac} also supports writing tactics that may generate an unknown number
of subgoals; \eg,~\fullref[bot]{fig:definetactic} sketches \ttt{inversion},
which for an existing theorem \ttt{H}, generates equalities that must also be
true based on the injectivity of constructors. For example, inverting \ttt{(=
  (S x) (S y))} produces \ttt{(= x y)}. The \ttt{inversion} tactic uses
\ttt{unify+prove}, which performs specialization by
unification~\cite{goguen2006}, computing either a series of equalities and
their proofs, or a proof of \ttt{False}. If the former, the current proof state
is extended with the new equalities; if the latter, the current goal is
immediately proved. Specifically, \ttt{inversion} unifies the indices of
\ttt{\ty[H]}, \ttt{H}'s type, with the indices of the result of each
constructor \ttt{C} of \ttt{\ty[H]}, creating a proof subnode for each
\ttt{C}. The result of \ttt{inversion} is an \ttt{elim} term for \ttt{H}, where
the bodies of the methods are the results of the given subgoals, referenced
with an implicit \ttt{\$pf} variable.

By implementing tactics as functions, \emph{tacticals}, \ie, tactic combinators,
are straightforward.
For example, here is \ttt{try}, which takes a sequence of tactics but backtracks if any of them fail:
\begin{alltt}\codefontsize
(define-tactical \defname{try}\moduname{ntac}
  [\pat{(\_ t \ooo)} \stx{(with-err-handler (\lm (e) \$ptz) ((compose (reverse (list t \ooo))) \$ptz))}])
\end{alltt}
It reverses its given tactics and applies them with with \ttt{compose},
which applies its rightmost argument first. If any of the tactics
errors, then \ttt{try} reverts to the original proof state
\ttt{\$ptz}.


Macros also enable more flexible control of the surface syntax, even allowing
hybrid tactic-tacticals. For example, here is a skeleton of an \ttt{induction} tactic with two cases:
\begin{alltt}\codefontsize
(define-tactic \defname{induction}\moduname{ntac}
  [\pat{(\_ H #:as ((x \ooo) \ooo))} \comm{\ooo}]
  [\pat{(\_ H [(C x ...) #:subgoal-is subg tactic \ooo] \ooo)} \comm{\ooo}])
\end{alltt}
The first is similar to systems like Coq, where the user supplies identifiers
that will bind the data constructor arguments in each case. Programmers,
however, can find it hard to read this kind of command. The second case
enables a slightly more ``declarative'' usage: each data
constructor case is named, and thus may appear in arbitrary order; the
subgoals are explicit and checked, making the script easier to follow; and
the tactics for each case are grouped, giving the proof more structure.

We can even equip user-defined tactics with features like interactivity:

\begin{minipage}[c]{0.5\textwidth}
\begin{alltt}\codefont
(define-tactic \defname{interactive}
 [\pat{\_} (print \$pt)
  (match (read-syntax)
   [\pat{(quit)} \$ptz]
   [\pat{a-tactic}\,\stx{(interactive\,(a-tactic}\,\$ptz\stx{))}])])
(ntac (\fatt (A : Type) (a : A) A) interactive)
\end{alltt}
\end{minipage}
\begin{minipage}[c]{0.49\textwidth}
\begin{alltt}\codefont
   goal 1 of 1: (\fatt (A : Type) (a : A) A)
   > (by-intros A a)
         ctx:  A : Type  a : A  (step #1)
               ---------------
   curr goal:  A
   > by-assumption
   Proof complete (2 steps)
   > (quit) \comm{script: \ttt{(intro A a) assumption}}
\end{alltt}
\end{minipage}
Specifically, the \ttt{interactive} tactic uses \ttt{print} to display
the proof state, then starts a read-eval-print-loop (REPL). The right side
shows an example interactive session. The REPL repeatedly reads in a command
and runs it; when it sees \ttt{quit}, it prints the complete proof script and
evaluates to the resulting proof term. We conjecture we could also
embed \modu{\ntac} with IDEs like emacs or DrRacket, perhaps using the
techniques of~\citet{korkut2018}, for even better interactivity.


\noindent\textbf{Programming with \Cur and \ntac} To demonstrate that one may
usefully program with the languages we create, we implemented a large test
suite. In particular, we spent one year using \modu{Cur} and \modu{\ntac}
to study the \emph{Software Foundations} curriculum (vol 1). Since it targets
novices, its examples cover a wide breadth of features and is thus a convenient
way to stress-test the flexibility of our macros-based approach to implementing
dependent types. This table summarizes our test suite:
\begin{tabular}{|lc|lc|lc|lc|}
\hline
\olly & 59 &
\modu{dep-lang} & 272 &
\modu{dep-ind-lang} & 2914 &
\modu{typed/video} & 721 \\
\ntac (sf vol 1) & 8045 &
sized types & 239 &
axioms & 98 & & \\
sugar & 58 &
patterns and def & 122 &
solver & 202 & \textbf{total}: \textasciitilde{13.3k} LoC & \\
\hline
\end{tabular}   



%% file: related.tex
There are many tutorials on \textbf{implementing dependent
types}~\cite{dep-lc-tutorial,simply-easier,simply-easy-ocaml, pi-forall,
without-sugar}. They typically start from scratch, however, \eg, they
manually manage type environments and rely on deBruijn indices to compute
$\alpha$-equality. They also often do not include practical features such as
user-defined inductive datatypes, nor are they easily extensible with sugar,
interactivity, or companion DSLs that programmers typically need to use with
their dependently typed language. In contrast, our macros-based approach
enables rapid creation of a core dependently typed language, and scales to a
full proof assistant.

\noindent\textbf{Extending proof assistants}, particularly via metaprogramming,
is an active area of research~\cite{devriese2013, christiansen2016,
  ebner2017:lean}. For some languages, however, this requires extending the
  core~\cite{brady2006b}. Other languages like Coq require writing extensions
  in a less integrated manner, \eg, programming plugins with OCaml and then
  linking it with other language binaries.  We present an alternative,
  linguistically integrated approach to extensibility, using the macro system
  inherited from the host language.

\noindent\textbf{New tactic languages} continue to make proof assistants easier to use~\cite{gonthier2010, gonthier2011, krebbers2017, malecha2016}. This
  suggests that (1) the ability to create a variety of tactic languages is
  critical, and (2) that linguistic support for creation of such DSLs would be
  well received. While we have yet to conduct a thorough comparison of all
  tactic languages and their implementations, we conjecture that our
  macros-based approach could accommodate many of them in a convenient
  manner. For example, there has been recent exploration of typed tactic
  languages Beluga~\cite{beluga}, Mtac~\cite{mtac}, and VeriML~\cite{veriml}.
  We conjecture that it would be straightforward to add a typed tactic language
  to \modu{Cur} using our macros-based approach.  This could be done either by
  utilizing \Turntwo, or using \modu{Cur}'s reflection API to use \modu{Cur} as
  it's own meta-language, following work in Lean~\cite{ebner2017:lean},
  Idris~\cite{christiansen2016}, Agda~\cite{norell2007}, or
  Coq~\cite{anand2018:typed-template-coq}.











\secsp
\section{Future Work}
In addition to exploring typed tactics and extensions like automation,
we will also continue adding features and improving various aspects of our
framework. For example, if \ttt{\defred} considered type
information in addition to a redex pattern, it would enable implementing
type-directed equality rules like \ttt{\texteta}. Another potential improvement
involves preventing abstraction leaks due to the interleaving of checking and
expansion.  For example, \modu{\Cur} and \modu{\ntac} must resugar during
interactive proofs to avoid exposing users to elaborated syntax. The current
approach is ad-hoc, however, recent advances~\cite{pombrio2015:resugaring}
could help. Another solution could be to implement a
\emph{domain-specific core target language}, thus avoiding resugaring
altogether. Such a feature would also improve the performance of type checking
by culling extraneous expansion steps. We are exploring such enhancements,
which possibly include changes to the macro expander itself, in order to
further advance type checking with macros.

%% file: conclusion.tex
To fully realize the benefits of dependent types, programmers should be able to
quickly develop dependently typed DSLs with the right power for a domain, and
rapidly iterate on new dependently typed language features. Further, such
languages should be easily extensible with new notation or companion DSLs that
may be required for practical use cases. We have demonstrated that a
macros-based approach to building dependently typed languages and features
satisfies this criteria.


%% file: acks.tex
\begin{acks}                            
  We acknowledge the support of the \grantsponsor{GS501100000023}{NSERC}{http://dx.doi.org/10.13039/501100000023} grant
  \grantnum{GS501100000023}{RGPIN-2019-04207}, and NSF grants \grantnum{GS100000001}{1823244} and \grantnum{GS100000001}{1518844}.
  Cette recherche a été financée par le \grantsponsor{GS501100000023}{CRSNG}{http://dx.doi.org/10.13039/501100000023}, numéro de référence
  \grantnum{GS501100000023}{RGPIN-2019-04207}.
\end{acks}

%% file: paper.bbl

\begin{thebibliography}{56}


\ifx \showCODEN    \undefined \def \showCODEN     #1{\unskip}     \fi
\ifx \showDOI      \undefined \def \showDOI       #1{#1}\fi
\ifx \showISBNx    \undefined \def \showISBNx     #1{\unskip}     \fi
\ifx \showISBNxiii \undefined \def \showISBNxiii  #1{\unskip}     \fi
\ifx \showISSN     \undefined \def \showISSN      #1{\unskip}     \fi
\ifx \showLCCN     \undefined \def \showLCCN      #1{\unskip}     \fi
\ifx \shownote     \undefined \def \shownote      #1{#1}          \fi
\ifx \showarticletitle \undefined \def \showarticletitle #1{#1}   \fi
\ifx \showURL      \undefined \def \showURL       {\relax}        \fi
\providecommand\bibfield[2]{#2}
\providecommand\bibinfo[2]{#2}
\providecommand\natexlab[1]{#1}
\providecommand\showeprint[2][]{arXiv:#2}

\bibitem[\protect\citeauthoryear{??}{rus}{2017}]%
        {rust2017b}
 \bibinfo{year}{2017}\natexlab{}.
\newblock \bibinfo{title}{{RFC}: The pi type trilogy}.
\newblock
\newblock
\urldef\tempurl%
\url{https://github.com/rust-lang/rfcs/issues/1930}
\showURL{%
\tempurl}


\bibitem[\protect\citeauthoryear{Abel}{Abel}{2010}]%
        {abel10sized}
\bibfield{author}{\bibinfo{person}{Andreas Abel}.}
  \bibinfo{year}{2010}\natexlab{}.
\newblock \showarticletitle{MiniAgda: Integrating Sized and Dependent Types}.
  In \bibinfo{booktitle}{\emph{PAR}} \emph{(\bibinfo{series}{EPTCS})},
  \bibfield{editor}{\bibinfo{person}{Ana Bove}, \bibinfo{person}{Ekaterina
  Komendantskaya}, {and} \bibinfo{person}{Milad Niqui}} (Eds.),
  Vol.~\bibinfo{volume}{43}. \bibinfo{pages}{14--28}.
\newblock
\urldef\tempurl%
\url{http://dblp.uni-trier.de/db/series/eptcs/eptcs43.html#abs-1012-4896}
\showURL{%
\tempurl}


\bibitem[\protect\citeauthoryear{Abel}{Abel}{2012}]%
        {abel12sized}
\bibfield{author}{\bibinfo{person}{Andreas Abel}.}
  \bibinfo{year}{2012}\natexlab{}.
\newblock \showarticletitle{Type-Based Termination, Inflationary Fixed-Points,
  and Mixed Inductive-Coinductive Types}. In
  \bibinfo{booktitle}{\emph{Proceedings 8th Workshop on Fixed Points in
  Computer Science, {FICS} 2012, Tallinn, Estonia, 24th March 2012.}}
  \bibinfo{pages}{1--11}.
\newblock
\urldef\tempurl%
\url{https://doi.org/10.4204/EPTCS.77.1}
\showDOI{\tempurl}


\bibitem[\protect\citeauthoryear{Altenkirch, Danielsson, L\"{o}h, and
  Oury}{Altenkirch et~al\mbox{.}}{2010}]%
        {without-sugar}
\bibfield{author}{\bibinfo{person}{Thorsten Altenkirch},
  \bibinfo{person}{Nils~Anders Danielsson}, \bibinfo{person}{Andres L\"{o}h},
  {and} \bibinfo{person}{Nicolas Oury}.} \bibinfo{year}{2010}\natexlab{}.
\newblock \showarticletitle{{$\Pi$}{$\Sigma$}: Dependent Types Without the
  Sugar}. In \bibinfo{booktitle}{\emph{Proceedings of the 10th International
  Conference on Functional and Logic Programming}}
  \emph{(\bibinfo{series}{FLOPS'10})}. \bibinfo{pages}{40--55}.
\newblock


\bibitem[\protect\citeauthoryear{Amin, Rompf, and Odersky}{Amin
  et~al\mbox{.}}{2014}]%
        {scalapathdep14}
\bibfield{author}{\bibinfo{person}{Nada Amin}, \bibinfo{person}{Tiark Rompf},
  {and} \bibinfo{person}{Martin Odersky}.} \bibinfo{year}{2014}\natexlab{}.
\newblock \showarticletitle{Foundations of Path-dependent Types}. In
  \bibinfo{booktitle}{\emph{Proceedings of the 2014 ACM International
  Conference on Object Oriented Programming Systems Languages \& Applications}}
  \emph{(\bibinfo{series}{OOPSLA '14})}. \bibinfo{publisher}{ACM},
  \bibinfo{address}{New York, NY, USA}, \bibinfo{pages}{233--249}.
\newblock
\showISBNx{978-1-4503-2585-1}
\urldef\tempurl%
\url{https://doi.org/10.1145/2660193.2660216}
\showDOI{\tempurl}


\bibitem[\protect\citeauthoryear{Anand, Boulier, Cohen, Sozeau, and
  Tabareau}{Anand et~al\mbox{.}}{2018}]%
        {anand2018:typed-template-coq}
\bibfield{author}{\bibinfo{person}{Abhishek Anand}, \bibinfo{person}{Simon
  Boulier}, \bibinfo{person}{Cyril Cohen}, \bibinfo{person}{Matthieu Sozeau},
  {and} \bibinfo{person}{Nicolas Tabareau}.} \bibinfo{year}{2018}\natexlab{}.
\newblock \showarticletitle{Towards Certified Meta-Programming with Typed
  Template-Coq}.
\newblock  (\bibinfo{year}{2018}).
\newblock
\urldef\tempurl%
\url{www.irif.fr/~sozeau/research/publications/drafts/Towards_Certified_Meta-Programming_with_Typed_Template-Coq.pdf}
\showURL{%
\tempurl}


\bibitem[\protect\citeauthoryear{Andersen, Chang, and Felleisen}{Andersen
  et~al\mbox{.}}{2017}]%
        {video}
\bibfield{author}{\bibinfo{person}{Leif Andersen}, \bibinfo{person}{Stephen
  Chang}, {and} \bibinfo{person}{Matthias Felleisen}.}
  \bibinfo{year}{2017}\natexlab{}.
\newblock \showarticletitle{Super 8 Languages for Making Movies (Functional
  Pearl)}.
\newblock \bibinfo{journal}{\emph{Proc. ACM Program. Lang.}}
  \bibinfo{volume}{1}, \bibinfo{number}{ICFP}, Article \bibinfo{articleno}{30}
  (\bibinfo{date}{Aug.} \bibinfo{year}{2017}), \bibinfo{numpages}{29}~pages.
\newblock
\showISSN{2475-1421}
\urldef\tempurl%
\url{https://doi.org/10.1145/3110274}
\showDOI{\tempurl}


\bibitem[\protect\citeauthoryear{Augustsson}{Augustsson}{2007}]%
        {simply-easier}
\bibfield{author}{\bibinfo{person}{Lennart Augustsson}.}
  \bibinfo{year}{2007}\natexlab{}.
\newblock \bibinfo{title}{Simpler, Easier!}
\newblock
\newblock
\urldef\tempurl%
\url{http://augustss.blogspot.ru/2007/10/simpler-easier-in-recent-paper-simply.html}
\showURL{%
\tempurl}


\bibitem[\protect\citeauthoryear{Bauer}{Bauer}{2012}]%
        {simply-easy-ocaml}
\bibfield{author}{\bibinfo{person}{Andrej Bauer}.}
  \bibinfo{year}{2012}\natexlab{}.
\newblock \bibinfo{title}{How to Implement Dependent Type Theory}.
\newblock
\newblock
\urldef\tempurl%
\url{http://math.andrej.com/2012/11/08/how-to-implement-dependent-type-theory-i/}
\showURL{%
\tempurl}


\bibitem[\protect\citeauthoryear{Blanchette, Kaliszyk, Paulson, and
  Urban}{Blanchette et~al\mbox{.}}{2016}]%
        {hammer2016}
\bibfield{author}{\bibinfo{person}{Jasmin~C. Blanchette},
  \bibinfo{person}{Cezary Kaliszyk}, \bibinfo{person}{Lawrence~C. Paulson},
  {and} \bibinfo{person}{Josef Urban}.} \bibinfo{year}{2016}\natexlab{}.
\newblock \showarticletitle{Hammering towards {QED}}.
\newblock \bibinfo{journal}{\emph{J. Formaliz. Reason.}} \bibinfo{volume}{9},
  \bibinfo{number}{1} (\bibinfo{year}{2016}), \bibinfo{pages}{101--148}.
\newblock
\showISSN{1972-5787}


\bibitem[\protect\citeauthoryear{Brady and Hammond}{Brady and Hammond}{2006}]%
        {brady2006b}
\bibfield{author}{\bibinfo{person}{Edwin Brady} {and} \bibinfo{person}{Kevin
  Hammond}.} \bibinfo{year}{2006}\natexlab{}.
\newblock \showarticletitle{Dependently Typed MetaProgramming}. In
  \bibinfo{booktitle}{\emph{7th Symposium on Trends in Functional
  Programming}}.
\newblock


\bibitem[\protect\citeauthoryear{Brady}{Brady}{2005}]%
        {brady-thesis}
\bibfield{author}{\bibinfo{person}{Edwin~C. Brady}.}
  \bibinfo{year}{2005}\natexlab{}.
\newblock \emph{\bibinfo{title}{Practical Implementation of a Dependently Typed
  Functional Programming Language}}.
\newblock \bibinfo{thesistype}{Ph.D. Dissertation}. \bibinfo{school}{University
  of Durham}.
\newblock


\bibitem[\protect\citeauthoryear{Chang, Knauth, and Greenman}{Chang
  et~al\mbox{.}}{2017}]%
        {macrotypes}
\bibfield{author}{\bibinfo{person}{Stephen Chang}, \bibinfo{person}{Alex
  Knauth}, {and} \bibinfo{person}{Ben Greenman}.}
  \bibinfo{year}{2017}\natexlab{}.
\newblock \showarticletitle{Type Systems As Macros}. In
  \bibinfo{booktitle}{\emph{Proceedings of the 44th ACM SIGPLAN Symposium on
  Principles of Programming Languages}}. \bibinfo{pages}{694--705}.
\newblock


\bibitem[\protect\citeauthoryear{Christiansen and Brady}{Christiansen and
  Brady}{2016}]%
        {christiansen2016}
\bibfield{author}{\bibinfo{person}{David Christiansen} {and}
  \bibinfo{person}{Edwin Brady}.} \bibinfo{year}{2016}\natexlab{}.
\newblock \showarticletitle{Elaborator Reflection: Extending Idris in Idris}.
  In \bibinfo{booktitle}{\emph{Proceedings of the 21st ACM SIGPLAN
  International Conference on Functional Programming}}
  \emph{(\bibinfo{series}{ICFP 2016})}. \bibinfo{publisher}{ACM},
  \bibinfo{address}{New York, NY, USA}, \bibinfo{pages}{284--297}.
\newblock
\showISBNx{978-1-4503-4219-3}
\urldef\tempurl%
\url{https://doi.org/10.1145/2951913.2951932}
\showDOI{\tempurl}


\bibitem[\protect\citeauthoryear{Cockx, Devriese, and Piessens}{Cockx
  et~al\mbox{.}}{2014}]%
        {cockx14}
\bibfield{author}{\bibinfo{person}{Jesper Cockx}, \bibinfo{person}{Dominique
  Devriese}, {and} \bibinfo{person}{Frank Piessens}.}
  \bibinfo{year}{2014}\natexlab{}.
\newblock \showarticletitle{Pattern Matching Without K}. In
  \bibinfo{booktitle}{\emph{Proceedings of the 19th ACM SIGPLAN International
  Conference on Functional Programming}} \emph{(\bibinfo{series}{ICFP '14})}.
  \bibinfo{publisher}{ACM}, \bibinfo{address}{New York, NY, USA},
  \bibinfo{pages}{257--268}.
\newblock
\showISBNx{978-1-4503-2873-9}
\urldef\tempurl%
\url{https://doi.org/10.1145/2628136.2628139}
\showDOI{\tempurl}


\bibitem[\protect\citeauthoryear{Coquand}{Coquand}{1992}]%
        {coquand1992}
\bibfield{author}{\bibinfo{person}{Thierry Coquand}.}
  \bibinfo{year}{1992}\natexlab{}.
\newblock \showarticletitle{Pattern Matching with Dependent Types}. In
  \bibinfo{booktitle}{\emph{Proceedings of the Workshop on Types for Proofs and
  Programs}}. \bibinfo{pages}{71--83}.
\newblock


\bibitem[\protect\citeauthoryear{Coquand and Huet}{Coquand and Huet}{1988}]%
        {coquand1988}
\bibfield{author}{\bibinfo{person}{Thierry Coquand} {and}
  \bibinfo{person}{G{\'{e}}rard~P. Huet}.} \bibinfo{year}{1988}\natexlab{}.
\newblock \showarticletitle{The Calculus of Constructions}.
\newblock \bibinfo{journal}{\emph{Inf. Comput.}} \bibinfo{volume}{76},
  \bibinfo{number}{2/3} (\bibinfo{year}{1988}), \bibinfo{pages}{95--120}.
\newblock
\urldef\tempurl%
\url{https://doi.org/10.1016/0890-5401(88)90005-3}
\showDOI{\tempurl}


\bibitem[\protect\citeauthoryear{Cremet, Garillot, Lenglet, and Odersky}{Cremet
  et~al\mbox{.}}{2006}]%
        {scalapathdep06}
\bibfield{author}{\bibinfo{person}{Vincent Cremet},
  \bibinfo{person}{Fran\c{c}ois Garillot}, \bibinfo{person}{Sergue\"{\i}
  Lenglet}, {and} \bibinfo{person}{Martin Odersky}.}
  \bibinfo{year}{2006}\natexlab{}.
\newblock \showarticletitle{A Core Calculus for Scala Type Checking}. In
  \bibinfo{booktitle}{\emph{Proceedings of the 31st International Conference on
  Mathematical Foundations of Computer Science}}
  \emph{(\bibinfo{series}{MFCS'06})}. \bibinfo{publisher}{Springer-Verlag},
  \bibinfo{address}{Berlin, Heidelberg}, \bibinfo{pages}{1--23}.
\newblock
\showISBNx{3-540-37791-3, 978-3-540-37791-7}
\urldef\tempurl%
\url{https://doi.org/10.1007/11821069_1}
\showDOI{\tempurl}


\bibitem[\protect\citeauthoryear{de~Bruijn}{de~Bruijn}{1991}]%
        {debruijn1991:telescope}
\bibfield{author}{\bibinfo{person}{N.G. de Bruijn}.}
  \bibinfo{year}{1991}\natexlab{}.
\newblock \showarticletitle{Telescopic Mappings in Typed Lambda-Calculus}.
\newblock \bibinfo{journal}{\emph{Information and Computation}}
  \bibinfo{volume}{91}, \bibinfo{number}{2} (\bibinfo{year}{1991}),
  \bibinfo{pages}{189--204}.
\newblock


\bibitem[\protect\citeauthoryear{Delahaye}{Delahaye}{2000}]%
        {delahaye2000}
\bibfield{author}{\bibinfo{person}{David Delahaye}.}
  \bibinfo{year}{2000}\natexlab{}.
\newblock \showarticletitle{A Tactic Language for the System Coq}. In
  \bibinfo{booktitle}{\emph{Proceedings of the 7th International Conference on
  Logic for Programming and Automated Reasoning}}
  \emph{(\bibinfo{series}{LPAR'00})}. \bibinfo{publisher}{Springer-Verlag},
  \bibinfo{address}{Berlin, Heidelberg}, \bibinfo{pages}{85--95}.
\newblock
\showISBNx{3-540-41285-9}
\urldef\tempurl%
\url{http://dl.acm.org/citation.cfm?id=1765236.1765246}
\showURL{%
\tempurl}


\bibitem[\protect\citeauthoryear{Devriese and Piessens}{Devriese and
  Piessens}{2013}]%
        {devriese2013}
\bibfield{author}{\bibinfo{person}{Dominique Devriese} {and}
  \bibinfo{person}{Frank Piessens}.} \bibinfo{year}{2013}\natexlab{}.
\newblock \showarticletitle{Typed Syntactic Meta-programming}. In
  \bibinfo{booktitle}{\emph{of the 18th ACM SIGPLAN International Conference on
  Functional Programming}} \emph{(\bibinfo{series}{ICFP 2013})}.
  \bibinfo{pages}{73–86}.
\newblock
\showISBNx{978-1-4503-2326-0}


\bibitem[\protect\citeauthoryear{Dybjer}{Dybjer}{1994}]%
        {Dybjer1994}
\bibfield{author}{\bibinfo{person}{Peter Dybjer}.}
  \bibinfo{year}{1994}\natexlab{}.
\newblock \showarticletitle{Inductive families}.
\newblock \bibinfo{journal}{\emph{Formal Aspects of Computing}}
  \bibinfo{volume}{6}, \bibinfo{number}{4} (\bibinfo{date}{01 Jul}
  \bibinfo{year}{1994}), \bibinfo{pages}{440--465}.
\newblock


\bibitem[\protect\citeauthoryear{Dybvig, Hieb, and Bruggeman}{Dybvig
  et~al\mbox{.}}{1992}]%
        {dybvig1992}
\bibfield{author}{\bibinfo{person}{R.~Kent Dybvig}, \bibinfo{person}{Robert
  Hieb}, {and} \bibinfo{person}{Carl Bruggeman}.}
  \bibinfo{year}{1992}\natexlab{}.
\newblock \showarticletitle{Syntactic Abstraction in Scheme}.
\newblock \bibinfo{journal}{\emph{Lisp Symb. Comput.}} \bibinfo{volume}{5},
  \bibinfo{number}{4} (\bibinfo{date}{Dec.} \bibinfo{year}{1992}),
  \bibinfo{pages}{295--326}.
\newblock
\showISSN{0892-4635}
\urldef\tempurl%
\url{https://doi.org/10.1007/BF01806308}
\showDOI{\tempurl}


\bibitem[\protect\citeauthoryear{Ebner, Ullrich, Roesch, Avigad, and
  de~Moura}{Ebner et~al\mbox{.}}{2017}]%
        {ebner2017:lean}
\bibfield{author}{\bibinfo{person}{Gabriel Ebner}, \bibinfo{person}{Sebastian
  Ullrich}, \bibinfo{person}{Jared Roesch}, \bibinfo{person}{Jeremy Avigad},
  {and} \bibinfo{person}{Leonardo de Moura}.} \bibinfo{year}{2017}\natexlab{}.
\newblock \showarticletitle{A metaprogramming framework for formal
  verification}.
\newblock \bibinfo{journal}{\emph{Proceedings of the {ACM} on Programming
  Languages ({PACMPL})}} \bibinfo{volume}{1}, \bibinfo{number}{{ICFP}}
  (\bibinfo{year}{2017}), \bibinfo{pages}{34:1--34:29}.
\newblock
\urldef\tempurl%
\url{https://doi.org/10.1145/3110278}
\showDOI{\tempurl}


\bibitem[\protect\citeauthoryear{Felleisen, Findler, Flatt, Krishnamurthi,
  Barzilay, McCarthy, and Tobin-Hochstadt}{Felleisen et~al\mbox{.}}{2015}]%
        {manifesto}
\bibfield{author}{\bibinfo{person}{Matthias Felleisen},
  \bibinfo{person}{Robert~Bruce Findler}, \bibinfo{person}{Matthew Flatt},
  \bibinfo{person}{Shriram Krishnamurthi}, \bibinfo{person}{Eli Barzilay},
  \bibinfo{person}{Jay McCarthy}, {and} \bibinfo{person}{Sam Tobin-Hochstadt}.}
  \bibinfo{year}{2015}\natexlab{}.
\newblock \showarticletitle{{The Racket Manifesto}}. In
  \bibinfo{booktitle}{\emph{1st Summit on Advances in Programming Languages
  (SNAPL 2015)}}. \bibinfo{pages}{113--128}.
\newblock


\bibitem[\protect\citeauthoryear{Felleisen, Findler, Flatt, Krishnamurthi,
  Barzilay, McCarthy, and Tobin-Hochstadt}{Felleisen et~al\mbox{.}}{2018}]%
        {lop}
\bibfield{author}{\bibinfo{person}{Matthias Felleisen},
  \bibinfo{person}{Robert~Bruce Findler}, \bibinfo{person}{Matthew Flatt},
  \bibinfo{person}{Shriram Krishnamurthi}, \bibinfo{person}{Eli Barzilay},
  \bibinfo{person}{Jay McCarthy}, {and} \bibinfo{person}{Sam Tobin-Hochstadt}.}
  \bibinfo{year}{2018}\natexlab{}.
\newblock \showarticletitle{A Programmable Programming Language}.
\newblock \bibinfo{journal}{\emph{Commun. ACM}} \bibinfo{volume}{61},
  \bibinfo{number}{3} (\bibinfo{date}{Feb.} \bibinfo{year}{2018}),
  \bibinfo{pages}{62--71}.
\newblock
\showISSN{0001-0782}
\urldef\tempurl%
\url{https://doi.org/10.1145/3127323}
\showDOI{\tempurl}


\bibitem[\protect\citeauthoryear{Flatt}{Flatt}{2002}]%
        {flatt2002}
\bibfield{author}{\bibinfo{person}{Matthew Flatt}.}
  \bibinfo{year}{2002}\natexlab{}.
\newblock \showarticletitle{Composable and Compilable Macros: You Want It
  when?}. In \bibinfo{booktitle}{\emph{Proceedings of the Seventh ACM SIGPLAN
  International Conference on Functional Programming}}
  \emph{(\bibinfo{series}{ICFP '02})}. \bibinfo{publisher}{ACM},
  \bibinfo{address}{New York, NY, USA}, \bibinfo{pages}{72--83}.
\newblock
\showISBNx{1-58113-487-8}
\urldef\tempurl%
\url{https://doi.org/10.1145/581478.581486}
\showDOI{\tempurl}


\bibitem[\protect\citeauthoryear{Flatt}{Flatt}{2016}]%
        {setsofscopes}
\bibfield{author}{\bibinfo{person}{Matthew Flatt}.}
  \bibinfo{year}{2016}\natexlab{}.
\newblock \showarticletitle{Binding As Sets of Scopes}. In
  \bibinfo{booktitle}{\emph{Proceedings of the 43rd Annual ACM SIGPLAN-SIGACT
  Symposium on Principles of Programming Languages}}.
  \bibinfo{pages}{705--717}.
\newblock


\bibitem[\protect\citeauthoryear{Flatt, Culpepper, Darais, and Findler}{Flatt
  et~al\mbox{.}}{2012}]%
        {fcdf:macrosworktogether}
\bibfield{author}{\bibinfo{person}{Matthew Flatt}, \bibinfo{person}{Ryan
  Culpepper}, \bibinfo{person}{David Darais}, {and}
  \bibinfo{person}{Robert~Bruce Findler}.} \bibinfo{year}{2012}\natexlab{}.
\newblock \showarticletitle{Macros That Work Together: Compile-time Bindings,
  Partial Expansion, and Definition Contexts}.
\newblock  \bibinfo{volume}{22}, \bibinfo{number}{2} (\bibinfo{date}{March}
  \bibinfo{year}{2012}), \bibinfo{pages}{181--216}.
\newblock
\showISSN{0956-7968}
\urldef\tempurl%
\url{https://doi.org/10.1017/S0956796812000093}
\showDOI{\tempurl}


\bibitem[\protect\citeauthoryear{Gim{\'{e}}nez}{Gim{\'{e}}nez}{1995}]%
        {gimenez1995}
\bibfield{author}{\bibinfo{person}{Eduardo Gim{\'{e}}nez}.}
  \bibinfo{year}{1995}\natexlab{}.
\newblock \showarticletitle{Codifying Guarded Definitions with Recursive
  Schemes}. In \bibinfo{booktitle}{\emph{International Workshop on Types for
  Proofs and Programs ({TYPES})}}.
\newblock
\showISBNx{3-540-60579-7}
\urldef\tempurl%
\url{https://doi.org/10.1007/3-540-60579-7_3}
\showDOI{\tempurl}


\bibitem[\protect\citeauthoryear{Goguen, McBride, and McKinna}{Goguen
  et~al\mbox{.}}{2006}]%
        {goguen2006}
\bibfield{author}{\bibinfo{person}{Healfdene Goguen}, \bibinfo{person}{Conor
  McBride}, {and} \bibinfo{person}{James McKinna}.}
  \bibinfo{year}{2006}\natexlab{}.
\newblock \showarticletitle{Eliminating Dependent Pattern Matching}. In
  \bibinfo{booktitle}{\emph{Algebra, Meaning, and Computation: Essays dedicated
  to Joseph A. Goguen on the Occasion of His 65th Birthday}},
  \bibfield{editor}{\bibinfo{person}{Kokichi Futatsugi},
  \bibinfo{person}{Jean-Pierre Jouannaud}, {and} \bibinfo{person}{Jos{\'e}
  Meseguer}} (Eds.). \bibinfo{publisher}{Springer Berlin Heidelberg},
  \bibinfo{address}{Berlin, Heidelberg}, \bibinfo{pages}{521--540}.
\newblock
\showISBNx{978-3-540-35464-2}
\urldef\tempurl%
\url{https://doi.org/10.1007/11780274_27}
\showDOI{\tempurl}


\bibitem[\protect\citeauthoryear{Gonthier and Mahboubi}{Gonthier and
  Mahboubi}{2010}]%
        {gonthier2010}
\bibfield{author}{\bibinfo{person}{Georges Gonthier} {and}
  \bibinfo{person}{Assia Mahboubi}.} \bibinfo{year}{2010}\natexlab{}.
\newblock \showarticletitle{{An introduction to small scale reflection in
  Coq}}.
\newblock \bibinfo{journal}{\emph{{Journal of Formalized Reasoning}}}
  \bibinfo{volume}{3}, \bibinfo{number}{2} (\bibinfo{year}{2010}),
  \bibinfo{pages}{95--152}.
\newblock
\urldef\tempurl%
\url{https://hal.inria.fr/inria-00515548}
\showURL{%
\tempurl}


\bibitem[\protect\citeauthoryear{Gonthier, Ziliani, Nanevski, and
  Dreyer}{Gonthier et~al\mbox{.}}{2011}]%
        {gonthier2011}
\bibfield{author}{\bibinfo{person}{Georges Gonthier}, \bibinfo{person}{Beta
  Ziliani}, \bibinfo{person}{Aleksandar Nanevski}, {and} \bibinfo{person}{Derek
  Dreyer}.} \bibinfo{year}{2011}\natexlab{}.
\newblock \showarticletitle{How to Make Ad Hoc Proof Automation Less Ad Hoc}.
  In \bibinfo{booktitle}{\emph{Proceedings of the 16th ACM SIGPLAN
  International Conference on Functional Programming}}
  \emph{(\bibinfo{series}{ICFP '11})}. \bibinfo{publisher}{ACM},
  \bibinfo{address}{New York, NY, USA}, \bibinfo{pages}{163--175}.
\newblock
\showISBNx{978-1-4503-0865-6}
\urldef\tempurl%
\url{https://doi.org/10.1145/2034773.2034798}
\showDOI{\tempurl}


\bibitem[\protect\citeauthoryear{Hughes, Pareto, and Sabry}{Hughes
  et~al\mbox{.}}{1996}]%
        {hughes96sizedtypes}
\bibfield{author}{\bibinfo{person}{John Hughes}, \bibinfo{person}{Lars Pareto},
  {and} \bibinfo{person}{Amr Sabry}.} \bibinfo{year}{1996}\natexlab{}.
\newblock \showarticletitle{Proving the Correctness of Reactive Systems Using
  Sized Types}. In \bibinfo{booktitle}{\emph{Proceedings of the 23rd ACM
  SIGPLAN-SIGACT Symposium on Principles of Programming Languages}}
  \emph{(\bibinfo{series}{POPL '96})}. \bibinfo{publisher}{ACM},
  \bibinfo{address}{New York, NY, USA}, \bibinfo{pages}{410--423}.
\newblock
\showISBNx{0-89791-769-3}
\urldef\tempurl%
\url{https://doi.org/10.1145/237721.240882}
\showDOI{\tempurl}


\bibitem[\protect\citeauthoryear{Korkut and Christiansen}{Korkut and
  Christiansen}{2018}]%
        {korkut2018}
\bibfield{author}{\bibinfo{person}{Joomy Korkut} {and} \bibinfo{person}{David
  Christiansen}.} \bibinfo{year}{2018}\natexlab{}.
\newblock \showarticletitle{Extensible Type-Directed Editing}. In
  \bibinfo{booktitle}{\emph{Proceedings of the Workshop on Type-Driven
  Development}}.
\newblock


\bibitem[\protect\citeauthoryear{Krebbers, Timany, and Birkedal}{Krebbers
  et~al\mbox{.}}{2017}]%
        {krebbers2017}
\bibfield{author}{\bibinfo{person}{Robbert Krebbers}, \bibinfo{person}{Amin
  Timany}, {and} \bibinfo{person}{Lars Birkedal}.}
  \bibinfo{year}{2017}\natexlab{}.
\newblock \showarticletitle{Interactive Proofs in Higher-order Concurrent
  Separation Logic}. In \bibinfo{booktitle}{\emph{Proceedings of the 44th ACM
  SIGPLAN Symposium on Principles of Programming Languages}}
  \emph{(\bibinfo{series}{POPL 2017})}. \bibinfo{publisher}{ACM},
  \bibinfo{address}{New York, NY, USA}, \bibinfo{pages}{205--217}.
\newblock
\showISBNx{978-1-4503-4660-3}
\urldef\tempurl%
\url{https://doi.org/10.1145/3009837.3009855}
\showDOI{\tempurl}


\bibitem[\protect\citeauthoryear{L{\"{o}}h, McBride, and Swierstra}{L{\"{o}}h
  et~al\mbox{.}}{2010}]%
        {dep-lc-tutorial}
\bibfield{author}{\bibinfo{person}{Andres L{\"{o}}h}, \bibinfo{person}{Conor
  McBride}, {and} \bibinfo{person}{Wouter Swierstra}.}
  \bibinfo{year}{2010}\natexlab{}.
\newblock \showarticletitle{A Tutorial Implementation of a Dependently Typed
  Lambda Calculus}.
\newblock \bibinfo{journal}{\emph{Fundam. Inform.}} \bibinfo{volume}{102},
  \bibinfo{number}{2} (\bibinfo{year}{2010}), \bibinfo{pages}{177--207}.
\newblock


\bibitem[\protect\citeauthoryear{Malecha and Bengtson}{Malecha and
  Bengtson}{2016}]%
        {malecha2016}
\bibfield{author}{\bibinfo{person}{Gregory Malecha} {and}
  \bibinfo{person}{Jesper Bengtson}.} \bibinfo{year}{2016}\natexlab{}.
\newblock \bibinfo{booktitle}{\emph{Programming Languages and Systems: 25th
  European Symposium on Programming, ESOP 2016, Held as Part of the European
  Joint Conferences on Theory and Practice of Software, ETAPS 2016, Eindhoven,
  The Netherlands, April 2-8, 2016, Proceedings}}.
\newblock \bibinfo{publisher}{Springer Berlin Heidelberg},
  \bibinfo{address}{Berlin, Heidelberg}, Chapter Extensible and Efficient
  Automation Through Reflective Tactics, \bibinfo{pages}{532--559}.
\newblock
\showISBNx{978-3-662-49498-1}
\urldef\tempurl%
\url{https://doi.org/10.1007/978-3-662-49498-1_21}
\showDOI{\tempurl}


\bibitem[\protect\citeauthoryear{Martelli and Montanari}{Martelli and
  Montanari}{1982}]%
        {martelli-montanari82}
\bibfield{author}{\bibinfo{person}{Alberto Martelli} {and} \bibinfo{person}{Ugo
  Montanari}.} \bibinfo{year}{1982}\natexlab{}.
\newblock \showarticletitle{An Efficient Unification Algorithm}.
\newblock \bibinfo{journal}{\emph{ACM Trans. Program. Lang. Syst.}}
  \bibinfo{volume}{4}, \bibinfo{number}{2} (\bibinfo{date}{April}
  \bibinfo{year}{1982}), \bibinfo{pages}{258--282}.
\newblock
\showISSN{0164-0925}
\urldef\tempurl%
\url{https://doi.org/10.1145/357162.357169}
\showDOI{\tempurl}


\bibitem[\protect\citeauthoryear{Martin-L{\"o}f}{Martin-L{\"o}f}{1975}]%
        {martin-loef1975}
\bibfield{author}{\bibinfo{person}{Per Martin-L{\"o}f}.}
  \bibinfo{year}{1975}\natexlab{}.
\newblock \showarticletitle{An intuitionistic theory of types: Predicative
  part}.
\newblock \bibinfo{journal}{\emph{Studies in Logic and the Foundations of
  Mathematics}}  \bibinfo{volume}{80} (\bibinfo{year}{1975}),
  \bibinfo{pages}{73--118}.
\newblock


\bibitem[\protect\citeauthoryear{McBride}{McBride}{2000}]%
        {mcbride2000:thesis}
\bibfield{author}{\bibinfo{person}{Conor McBride}.}
  \bibinfo{year}{2000}\natexlab{}.
\newblock \emph{\bibinfo{title}{Dependently Typed Functional Programs and Their
  Proofs}}.
\newblock \bibinfo{thesistype}{Ph.D. Dissertation}. \bibinfo{school}{University
  of Edinburgh, {UK}}.
\newblock
\urldef\tempurl%
\url{http://hdl.handle.net/1842/374}
\showURL{%
\tempurl}


\bibitem[\protect\citeauthoryear{Nordstr\"{o}m, Petersson, and
  Smith}{Nordstr\"{o}m et~al\mbox{.}}{1990}]%
        {mlttbook}
\bibfield{author}{\bibinfo{person}{Bengt Nordstr\"{o}m}, \bibinfo{person}{Kent
  Petersson}, {and} \bibinfo{person}{Jan~M. Smith}.}
  \bibinfo{year}{1990}\natexlab{}.
\newblock \bibinfo{booktitle}{\emph{Programming in Martin-L{\"o}f's Type
  Theory: An Introduction}}.
\newblock \bibinfo{publisher}{Clarendon Press}, \bibinfo{address}{New York, NY,
  USA}.
\newblock
\showISBNx{0-19-853814-6}


\bibitem[\protect\citeauthoryear{Norell}{Norell}{2007}]%
        {norell2007}
\bibfield{author}{\bibinfo{person}{Ulf Norell}.}
  \bibinfo{year}{2007}\natexlab{}.
\newblock \emph{\bibinfo{title}{Towards a Practical Programming Language Based
  on Dependent Type Theory}}.
\newblock \bibinfo{thesistype}{Ph.D. Dissertation}. \bibinfo{school}{Chalmers
  University of Technology}.
\newblock
\showISBNx{978-91-7291-996-9}
\urldef\tempurl%
\url{http://www.cse.chalmers.se/~ulfn/papers/thesis.pdf}
\showURL{%
\tempurl}


\bibitem[\protect\citeauthoryear{Pientka}{Pientka}{2008}]%
        {beluga}
\bibfield{author}{\bibinfo{person}{Brigitte Pientka}.}
  \bibinfo{year}{2008}\natexlab{}.
\newblock \showarticletitle{A Type-theoretic Foundation for Programming with
  Higher-order Abstract Syntax and First-class Substitutions}. In
  \bibinfo{booktitle}{\emph{Proceedings of the 35th Annual ACM SIGPLAN-SIGACT
  Symposium on Principles of Programming Languages}}
  \emph{(\bibinfo{series}{POPL '08})}. \bibinfo{pages}{371--382}.
\newblock
\showISBNx{978-1-59593-689-9}


\bibitem[\protect\citeauthoryear{Pierce, de~Amorim, Casinghino, Gaboardi,
  Greenberg, Hri\c{t}cu, Sj\"{o}berg, and Yorgey}{Pierce et~al\mbox{.}}{2018}]%
        {sf1}
\bibfield{author}{\bibinfo{person}{Benjamin~C. Pierce},
  \bibinfo{person}{Arthur~Azevedo de Amorim}, \bibinfo{person}{Chris
  Casinghino}, \bibinfo{person}{Marco Gaboardi}, \bibinfo{person}{Michael
  Greenberg}, \bibinfo{person}{C\v{a}t\v{a}lin Hri\c{t}cu},
  \bibinfo{person}{Vilhelm Sj\"{o}berg}, {and} \bibinfo{person}{Brent Yorgey}.}
  \bibinfo{year}{2018}\natexlab{}.
\newblock \bibinfo{booktitle}{\emph{Logical Foundations}}.
\newblock \bibinfo{publisher}{Electronic textbook}.
\newblock


\bibitem[\protect\citeauthoryear{Pierce and Turner}{Pierce and Turner}{1998}]%
        {bidir-popl1998}
\bibfield{author}{\bibinfo{person}{Benjamin~C. Pierce} {and}
  \bibinfo{person}{David~N. Turner}.} \bibinfo{year}{1998}\natexlab{}.
\newblock \showarticletitle{Local Type Inference}. In
  \bibinfo{booktitle}{\emph{Proceedings of the 25th {ACM} {SIGPLAN--SIGACT}
  {S}ymposium on {P}rinciples of {P}rogramming {L}anguages}}.
  \bibinfo{pages}{252--265}.
\newblock


\bibitem[\protect\citeauthoryear{Pombrio and Krishnamurthi}{Pombrio and
  Krishnamurthi}{2015}]%
        {pombrio2015:resugaring}
\bibfield{author}{\bibinfo{person}{Justin Pombrio} {and}
  \bibinfo{person}{Shriram Krishnamurthi}.} \bibinfo{year}{2015}\natexlab{}.
\newblock \showarticletitle{Hygienic Resugaring of Compositional Desugaring}.
  In \bibinfo{booktitle}{\emph{Proceedings of the 20th ACM SIGPLAN
  International Conference on Functional Programming}}
  \emph{(\bibinfo{series}{ICFP 2015})}. \bibinfo{publisher}{ACM},
  \bibinfo{address}{New York, NY, USA}, \bibinfo{pages}{75--87}.
\newblock
\showISBNx{978-1-4503-3669-7}
\urldef\tempurl%
\url{https://doi.org/10.1145/2784731.2784755}
\showDOI{\tempurl}


\bibitem[\protect\citeauthoryear{Sewell, Nardelli, Owens, Peskine, Ridge,
  Sarkar, and Strni\v{s}a}{Sewell et~al\mbox{.}}{2007}]%
        {sewell:2007}
\bibfield{author}{\bibinfo{person}{Peter Sewell},
  \bibinfo{person}{Francesco~Zappa Nardelli}, \bibinfo{person}{Scott Owens},
  \bibinfo{person}{Gilles Peskine}, \bibinfo{person}{Thomas Ridge},
  \bibinfo{person}{Susmit Sarkar}, {and} \bibinfo{person}{Rok Strni\v{s}a}.}
  \bibinfo{year}{2007}\natexlab{}.
\newblock \showarticletitle{Ott: Effective Tool Support for the Working
  Semanticist}. In \bibinfo{booktitle}{\emph{of the 12th ACM SIGPLAN
  International Conference on Functional Programming}}
  \emph{(\bibinfo{series}{ICFP 2007})}. \bibinfo{publisher}{ACM},
  \bibinfo{address}{New York, NY, USA}, \bibinfo{pages}{1--12}.
\newblock
\showISBNx{978-1-59593-815-2}
\urldef\tempurl%
\url{https://doi.org/10.1145/1291151.1291155}
\showDOI{\tempurl}


\bibitem[\protect\citeauthoryear{Stampoulis and Shao}{Stampoulis and
  Shao}{2010}]%
        {veriml}
\bibfield{author}{\bibinfo{person}{Antonis Stampoulis} {and}
  \bibinfo{person}{Zhong Shao}.} \bibinfo{year}{2010}\natexlab{}.
\newblock \showarticletitle{VeriML: Typed Computation of Logical Terms Inside a
  Language with Effects}. In \bibinfo{booktitle}{\emph{of the 15th ACM SIGPLAN
  International Conference on Functional Programming}}
  \emph{(\bibinfo{series}{ICFP 2010})}. \bibinfo{pages}{333--344}.
\newblock
\showISBNx{978-1-60558-794-3}


\bibitem[\protect\citeauthoryear{Torlak and Bodik}{Torlak and Bodik}{2014}]%
        {rosette}
\bibfield{author}{\bibinfo{person}{Emina Torlak} {and}
  \bibinfo{person}{Rastislav Bodik}.} \bibinfo{year}{2014}\natexlab{}.
\newblock \showarticletitle{A Lightweight Symbolic Virtual Machine for
  Solver-aided Host Languages}. In \bibinfo{booktitle}{\emph{Proceedings of the
  35th ACM SIGPLAN Conference on Programming Language Design and
  Implementation}}. \bibinfo{pages}{530--541}.
\newblock


\bibitem[\protect\citeauthoryear{Weirich}{Weirich}{2014}]%
        {pi-forall}
\bibfield{author}{\bibinfo{person}{Stephanie Weirich}.}
  \bibinfo{year}{2014}\natexlab{}.
\newblock \bibinfo{title}{Pi Forall: notes from OPLSS}.
\newblock
\newblock
\urldef\tempurl%
\url{https://github.com/sweirich/pi-forall}
\showURL{%
\tempurl}


\bibitem[\protect\citeauthoryear{Weirich, Voizard, de~Amorim, and
  Eisenberg}{Weirich et~al\mbox{.}}{2017}]%
        {weirich2017}
\bibfield{author}{\bibinfo{person}{Stephanie Weirich}, \bibinfo{person}{Antoine
  Voizard}, \bibinfo{person}{Pedro Henrique~Avezedo de Amorim}, {and}
  \bibinfo{person}{Richard~A. Eisenberg}.} \bibinfo{year}{2017}\natexlab{}.
\newblock \showarticletitle{A Specification for Dependent Types in Haskell}.
\newblock \bibinfo{journal}{\emph{Proceedings of the {ACM} on Programming
  Languages ({PACMPL})}} \bibinfo{volume}{1}, \bibinfo{number}{{ICFP}}
  (\bibinfo{date}{Aug.} \bibinfo{year}{2017}).
\newblock
\urldef\tempurl%
\url{https://doi.org/10.1145/3110275}
\showDOI{\tempurl}


\bibitem[\protect\citeauthoryear{Xi}{Xi}{2007}]%
        {dependentml}
\bibfield{author}{\bibinfo{person}{Hongwei Xi}.}
  \bibinfo{year}{2007}\natexlab{}.
\newblock \showarticletitle{Dependent ML An approach to practical programming
  with dependent types}.
\newblock \bibinfo{journal}{\emph{Journal of Functional Programming}}
  \bibinfo{volume}{17}, \bibinfo{number}{2} (\bibinfo{year}{2007}),
  \bibinfo{pages}{215–286}.
\newblock
\urldef\tempurl%
\url{https://doi.org/10.1017/S0956796806006216}
\showDOI{\tempurl}


\bibitem[\protect\citeauthoryear{Yorgey, Weirich, Cretin, Jones, Vytiniotis,
  and Magalh{\~{a}}es}{Yorgey et~al\mbox{.}}{2012}]%
        {yorgey2012}
\bibfield{author}{\bibinfo{person}{Brent~A. Yorgey}, \bibinfo{person}{Stephanie
  Weirich}, \bibinfo{person}{Julien Cretin}, \bibinfo{person}{Simon L.~Peyton
  Jones}, \bibinfo{person}{Dimitrios Vytiniotis}, {and}
  \bibinfo{person}{Jos{\'{e}}~Pedro Magalh{\~{a}}es}.}
  \bibinfo{year}{2012}\natexlab{}.
\newblock \showarticletitle{Giving Haskell a promotion}. In
  \bibinfo{booktitle}{\emph{Types in Language Design and Implementation
  ({TLDI})}}.
\newblock
\urldef\tempurl%
\url{https://doi.org/10.1145/2103786.2103795}
\showDOI{\tempurl}


\bibitem[\protect\citeauthoryear{Ziliani, Dreyer, Krishnaswami, Nanevski, and
  Vafeiadis}{Ziliani et~al\mbox{.}}{2013}]%
        {mtac}
\bibfield{author}{\bibinfo{person}{Beta Ziliani}, \bibinfo{person}{Derek
  Dreyer}, \bibinfo{person}{Neelakantan~R Krishnaswami},
  \bibinfo{person}{Aleksandar Nanevski}, {and} \bibinfo{person}{Viktor
  Vafeiadis}.} \bibinfo{year}{2013}\natexlab{}.
\newblock \showarticletitle{Mtac: A Monad for Typed Tactic Programming in Coq}.
  In \bibinfo{booktitle}{\emph{Proceedings of the 18th ACM SIGPLAN
  International Conference on Functional Programming}}
  \emph{(\bibinfo{series}{ICFP 2013})}. \bibinfo{publisher}{ACM},
  \bibinfo{address}{New York, NY, USA}, \bibinfo{pages}{87--100}.
\newblock
\showISBNx{978-1-4503-2326-0}
\urldef\tempurl%
\url{https://doi.org/10.1145/2500365.2500579}
\showDOI{\tempurl}


\bibitem[\protect\citeauthoryear{Zinzindohou{\'{e}}, Bhargavan, Protzenko, and
  Beurdouche}{Zinzindohou{\'{e}} et~al\mbox{.}}{2017}]%
        {zinzindohoue2017}
\bibfield{author}{\bibinfo{person}{Jean~Karim Zinzindohou{\'{e}}},
  \bibinfo{person}{Karthikeyan Bhargavan}, \bibinfo{person}{Jonathan
  Protzenko}, {and} \bibinfo{person}{Benjamin Beurdouche}.}
  \bibinfo{year}{2017}\natexlab{}.
\newblock \showarticletitle{HACL*: {A} Verified Modern Cryptographic Library}.
  In \bibinfo{booktitle}{\emph{Conference on Computer and Communications
  Security, ({CCS})}}.
\newblock
\urldef\tempurl%
\url{https://doi.org/10.1145/3133956.3134043}
\showDOI{\tempurl}


\end{thebibliography}
